\documentclass[notitlepage,nofootinbib,preprintnumbers,onecolumn,amssymb,superscriptaddress]{revtex4-2}
\usepackage{amsfonts,amssymb,mathtools,graphicx,color,bm}
\definecolor{ultramarine}{rgb}{0.07, 0.04, 0.56}
\definecolor{cadmiumgreen}{rgb}{0.0, 0.42, 0.24}
\definecolor{indigo(dye)}{rgb}{0.0, 0.25, 0.42}
\usepackage[linktocpage=true]{hyperref}
\hypersetup{
colorlinks=true,
citecolor=ultramarine,
linkcolor=cadmiumgreen,
urlcolor=indigo(dye),
}

\usepackage{autobreak,braket}

\newcommand{\fr}[2]{\frac{#1}{#2}}

\newcommand{\ti}{\tilde}

\newcommand{\be}{\begin{equation}}  
\newcommand{\ee}{\end{equation}}
\newcommand{\beq}{\begin{eqnarray}}  
\newcommand{\eeq}{\end{eqnarray}}
\newcommand{\bem}{\begin{bmatrix}}
\newcommand{\eem}{\end{bmatrix}}

\newcommand{\mn}{{\mu \nu}}

\begin{document}

\preprint{YITP-24-44, IPMU24-0014, RIKEN-iTHEMS-Report-24}

\title{Black holes, multiple propagation speeds, and energy extraction}

\author{Vitor Cardoso}
\affiliation{Niels Bohr International Academy, Niels Bohr Institute, Blegdamsvej 17, 2100 Copenhagen, Denmark}
\affiliation{CENTRA, Departamento de F\'{\i}sica, Instituto Superior T\'ecnico -- IST, Universidade de Lisboa -- UL,
Avenida Rovisco Pais 1, 1049-001 Lisboa, Portugal}

\author{Shinji Mukohyama}
\affiliation{Center for Gravitational Physics and Quantum Information, Yukawa Institute for Theoretical Physics, Kyoto University, 606-8502, Kyoto, Japan}
\affiliation{Kavli Institute for the Physics and Mathematics of the Universe (WPI), The University of Tokyo Institutes for Advanced Study, The University of Tokyo, Kashiwa, Chiba 277-8583, Japan}

\author{Naritaka Oshita}
\affiliation{Center for Gravitational Physics and Quantum Information, Yukawa Institute for Theoretical Physics, Kyoto University, 606-8502, Kyoto, Japan}
\affiliation{The Hakubi Center for Advanced Research, Kyoto University,
Yoshida Ushinomiyacho, Sakyo-ku, Kyoto 606-8501, Japan}
\affiliation{RIKEN iTHEMS, Wako, Saitama, 351-0198, Japan}

\author{Kazufumi Takahashi}
\affiliation{Center for Gravitational Physics and Quantum Information, Yukawa Institute for Theoretical Physics, Kyoto University, 606-8502, Kyoto, Japan}

\begin{abstract}
The Standard Model of particle physics predicts the speed of light to be a universal speed of propagation of massless carriers. However, other possibilities exist---including Lorentz-violating theories---where different fundamental fields travel at different speeds. Black holes are interesting probes of such physics, as distinct fields would probe different horizons. Here, we build an exact spacetime for two interacting scalar fields which have different propagation speeds. One of these fields is able to probe the black hole interior of the other, giving rise to energy extraction from the black hole and a characteristic late-time relaxation. Our results provide further stimulus to the search for extra degrees of freedom, black hole instability, and extra ringdown modes in gravitational-wave events.
\end{abstract}

\maketitle

\section{Introduction}\label{sec:intro}
General relativity is an extremely successful description of the gravitational interaction, which passed numerous tests, ranging over different orders of magnitude in length scale and field strength~\cite{Will:2014kxa}.
The advent of gravitational-wave astronomy~\cite{LIGOScientific:2016aoc} has opened the way for unprecedented tests of the strong-field regime, including new tests of Einstein's theory and the strengthening of the black hole (BH) paradigm~\cite{Berti:2015itd,Yunes:2016jcc,Barack:2018yly,Cardoso:2019rvt,LIGOScientific:2019fpa,LIGOScientific:2020tif,LIGOScientific:2021sio}.

BHs are a bizarre solution of the mathematical equations of general relativity, which nature seems to abide to~\cite{Hawking:1987en,Misner:1973prb,Chandrasekhar:1985kt,Cardoso:2019rvt}. Their role in fundamental physics is highlighted by two powerful results. The first concerns BH uniqueness, the most general vacuum BH solution belongs to the Kerr family~\cite{Carter:1971zc,Robinson:1975bv,Chrusciel:2012jk}. The second is that BH interiors harbor the failure of the underlying theory or setup from which the very notion of BHs arises~\cite{Penrose:1964wq,Penrose:1969pc}.

\begin{figure}[h]
\centering
\includegraphics[width=0.45\linewidth]{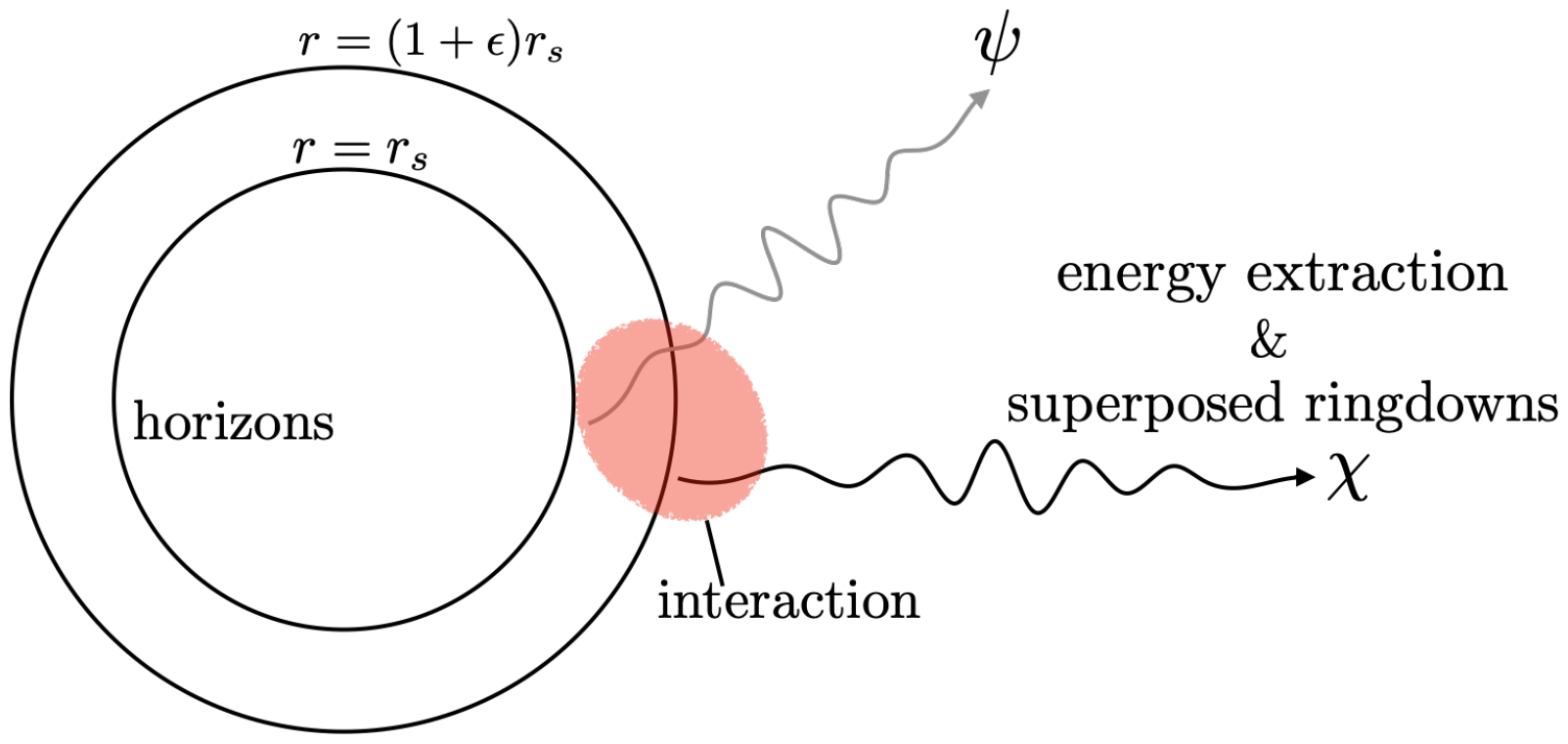}
\caption{Schematic picture of how we could probe the doubled horizons due to the difference in the speed of propagation of fields, $\psi$ and $\chi$. The parameter~$\epsilon$ controls the speed of the propagation of $\chi$. The ringdown emission and energy extraction from a nonspinning BH are discussed in Secs.~\ref{sec_QNM} and \ref{sec_SR}, respectively.}
\label{pic_schematic}
\end{figure}
In the Standard Model of particle physics, together with general relativity, massless fields travel at the speed of light. Hence, spacetime horizons are common to all interactions. However, this need not be the case. One can consider, for example, Lorentz-violating theories where propagation speeds are different and hence different interactions-carriers travel at different speeds.\footnote{In this context, the Lorentz violation could be either explicit or spontaneous,
depending on whether the vacuum expectation value is forced to be
nonvanishing at the level of the underlining theory (e.g.,
the khronometric theory~\cite{Blas:2010hb}) or it happens to be nonvanishing as a solution
to the underlining theory (e.g., ghost condensate~\cite{ArkaniHamed:2003uy}).} In such setups, one field, say $\psi$, is able to probe the region inside the horizon of another field~$\chi$, possibly carrying important information about BH interiors. We refer to Fig.~\ref{pic_schematic}.

For example, in the context of ghost condensation~\cite{ArkaniHamed:2003uy}, a scalar field~$\phi$ with a timelike gradient~$\phi_{\mu}\coloneqq \partial_{\mu}\phi$ provides a preferred time slicing and, as a result, spontaneously breaks Lorentz invariance. A spherically symmetric BH solution in ghost condensation was constructed in Ref.~\cite{Mukohyama:2005rw} by taking into account the effect of a higher-derivative term, which is present in the theory of ghost condensation (and which was later dubbed scordatura~\cite{Motohashi:2019ymr}), to the stealth Schwarzschild solution found for the first time also in Ref.~\cite{Mukohyama:2005rw}.
(See also Ref.~\cite{DeFelice:2022qaz} for an extension to more generic scalar-tensor theories with higher derivatives.)
For any practical purposes at astrophysical scales, the metric is well approximated by the Schwarzschild geometry, and the gradient~$\phi^{\mu}\coloneqq g^{\mu\nu}\phi_{\nu}$ is tangent to a congruence of geodesics which is regular on the BH horizon. As we shall see in the following section, coupling between $\phi^{\mu}$ and the derivative of other scalar fields can change the propagation speeds of the latter fields and thus provides a concrete setup for the situation depicted in Fig.~\ref{pic_schematic}.

An important effect of having different horizons for different fields is the possibility of energy extraction: inside $\chi$'s horizon, it can have negative energy, which may, via coupling between $\chi$ and $\psi$, be created together with positive energy of $\psi$. While the former (negative energy) remains inside the horizon, the latter (positive energy) can be extracted to the BH exterior~\cite{Eling:2007qd,Brito:2015oca}. No concrete realization of this setup is known, nor is there any evidence that energy extraction indeed occurs. In addition, even if no energy extraction occurs, the probing of the horizon interior may affect the dynamical behavior in the BH exterior, i.e., ringdown or the excitation of quasinormal modes (QNMs). Unfortunately, nothing is known about possible characteristic signatures.
The goal of this work is to fill the gap. 
We discuss and study a simple example of a theory with two interacting scalar fields, which probe different spacetime geometries, where all of the above is realized.

The rest of this paper is organized as follows.
In Sec.~\ref{sec:model}, we introduce an example of a theory where two scalar fields couple to two different effective geometries and have different propagation speeds.
In Sec.~\ref{numerics}, we provide details of the procedure to solve numerically the relevant equations on a fixed background. In Sec.~\ref{sec:results}, we provide a number of important benchmark tests. We also show two of our main results: {\bf i.}~the late-time relaxation contains imprints of {\it both} horizons, modulating the ringdown phase, with exciting observational consequences, and {\bf ii.}~energy extraction occurs in our setup, via scattering of wave packets.
Finally, we draw our conclusions in Sec.~\ref{sec:conc}.

\section{Framework}\label{sec:model}

We want to investigate a setup where different fields propagate at different speeds, by coupling to different effective spacetime geometries. In particular, we focus on two scalar fields~$\psi$ and $\chi$ that propagate on metrics~$g_{\mu\nu}$ and $\tilde{g}_{\mu\nu}$, respectively, and that interact with each other. For that, we consider a scalar field~$\psi$ coupled to a metric~$g_{\mu\nu}$, and a scalar~$\chi$ coupled to a disformally transformed version of it, $\tilde{g}_{\mu\nu}$.

Our toy model that realizes such a setup is defined by the action,
\beq
I&=&\int d^4x \sqrt{-g} 
\biggl( -\frac{1}{2} g^{\mu\nu}\partial_{\mu}\psi\partial_{\nu}\psi-\frac{1}{2}g^{\mu\nu}\partial_{\mu}\chi\partial_{\nu}\chi
+\frac{\epsilon}{2}(\phi^{\mu}\partial_{\mu}\chi)^2\nonumber\\
&-&\frac{1}{2}(\nabla_{\mu}\phi^{\mu})^2(\epsilon_{11}\chi^2 + 2\epsilon_{12}\chi\psi + \epsilon_{22}\psi^2)\biggr) \,,
\label{eqn:action1}
\eeq
where $\phi^{\mu} = g^{\mu\nu}\partial_{\nu}\phi$ with $\phi$ being a scalar field that satisfies $g^{\mu\nu}\partial_{\mu}\phi\partial_{\nu}\phi=-1$, and $\epsilon$ and $\epsilon_{11,12,22}$ are dimensionless constants.
Note that the $Z_2$ symmetry of $\phi$ is respected and the interaction between the fields~$\chi$ and $\psi$ is turned off at infinity if $\phi^{\mu}$ is chosen to approach the asymptotic timelike Killing vector at infinity.
This action can be rewritten as
\beq
I&=&\int d^4x \sqrt{-g} \biggl(  -\frac{1}{2} g^{\mu\nu}\partial_{\mu}\psi\partial_{\nu}\psi 
-\frac{1}{2}\tilde{g}^{\mu\nu}\partial_{\mu}\chi\partial_{\nu}\chi \nonumber\\
&-&\frac{1}{2}(\nabla_{\mu}\phi^{\mu})^2(\epsilon_{11}\chi^2 +2\epsilon_{12}\chi\psi+\epsilon_{22}\psi^2)\biggr) \,.
\label{eqn:action2}
\eeq
Here, the inverse disformal metric has been defined as $\tilde{g}^{\mu\nu}\coloneqq g^{\mu\nu}-\epsilon\phi^\mu\phi^\nu$, which corresponds to the following disformal metric:
\begin{align}
\tilde{g}_{\mu\nu}=g_{\mu\nu}+\frac{\epsilon}{1+\epsilon}\phi_\mu\phi_\nu\,.
\end{align}
As can be seen from Eq.~\eqref{eqn:action2}, the effective metric governing the dynamics of $\chi$ is the disformal metric~$\tilde{g}_{\mu\nu}$, which differs from the metric~$g_{\mu\nu}$ governing the scalar~$\psi$ when $\epsilon\ne 0$.
Therefore, in general the horizon of $\chi$ is different from that of $\psi$.

The model~\eqref{eqn:action2} is symmetric under the exchange of $\psi$ and $\chi$ in the sense that the action is invariant under the replacements
\begin{align}
&\psi \to (1+\epsilon)^{1/4}\chi\,, \quad \chi \to (1+\epsilon)^{1/4}\psi\,, \nonumber\\
& g_{\mu\nu} \to \tilde{g}_{\mu\nu}\,, \quad \phi \to (1+\epsilon)^{-1/2}\phi\,, \nonumber\\
& \epsilon \to -\frac{\epsilon}{1+\epsilon}\,, \quad \epsilon_{ij} \to (1+\epsilon)^{-1}\epsilon_{ji}\,, 
\label{eqn:psi-chi-duality}
\end{align} 
provided that $\epsilon>-1$, where we have used the fact that $\sqrt{-\tilde{g}}=(1+\epsilon)^{-1/2}\sqrt{-g}$ and that $\tilde{g}_{\mu\nu}$ and $\nabla_{\mu}\phi^{\mu}$ transform under \eqref{eqn:psi-chi-duality} as
\begin{align}
 \tilde{g}_{\mu\nu} \to g_{\mu\nu}\,, \qquad \nabla_{\mu}\phi^{\mu} \to (1+\epsilon)^{1/2}\nabla_{\mu}\phi^{\mu}\,.
\end{align} 
Under the duality transformation~\eqref{eqn:psi-chi-duality}, $\epsilon\in (-1,0)$ is mapped to $\epsilon\in (0,\infty)$, and vice versa. Therefore, in what follows, we assume $\epsilon>0$ without loss of generality. Note that the condition~$g^{\mu\nu}\partial_{\mu}\phi\partial_{\nu}\phi=-1$ is preserved under the above replacements.

For concreteness, consider a spherically symmetric, static spacetime,
\beq
g_{\mu\nu}dx^{\mu}dx^{\nu}&=& -Adt^2 + \frac{dr^2}{B}+ r^2 d\Omega^2 \nonumber \\
&=& -d\tau^2 + \left(1-A\right)d\rho^2 + r^2d\Omega^2\,,
\eeq
where $d\Omega^2$ is the metric of the unit $2$-sphere, $A=A(r)$, and $B=B(r)$.
Here, $\tau$ and $\rho$ represent Gaussian-normal (or Lema{\^i}tre) coordinates defined by
\begin{equation}
d\tau = dt + \sqrt{\frac{1-A}{AB}} dr\,, \qquad
d\rho = dt + \frac{dr}{\sqrt{AB(1-A)}}\,.
\label{tau_rho_coordinates}
\end{equation}
Note that $\partial_{\rho}r = -\partial_{\tau}r = \sqrt{B(1-A)/A}$.
The condition~$g^{\mu\nu}\partial_{\mu}\phi\partial_{\nu}\phi=-1$ can be realized if we choose
\begin{equation}
 \phi=\tau\,, 
\end{equation}
which is precisely the typical scalar field profile for the stealth Schwarzschild solution in the Gaussian-normal coordinate system~\cite{Mukohyama:2005rw,Khoury:2020aya,Takahashi:2021bml}. The corresponding disformally transformed metric is 
\begin{equation}
\tilde{g}_{\mu\nu}dx^{\mu}dx^{\nu} = -\frac{d\tau^2}{1+\epsilon}
+ \left(1-A\right)d\rho^2 + r^2d\Omega^2\,.
\label{scaled_metric}
\end{equation}
Note also that we assume $\epsilon>-1$ so that the coordinate~$\tau$ is timelike with respect to not only $g_{\mu\nu}$ but also $\tilde{g}_{\mu\nu}$.

In what follows, we specialize to the only vacuum static solution of general relativity, i.e., the Schwarzschild background where
\begin{align}
A=B=1-\fr{r_s}{r}\eqqcolon f\,.
\end{align}
We find in the standard Schwarzschild coordinates
\beq
g_{\mu\nu}dx^{\mu}dx^{\nu} &=& -fdt^2 + \frac{dr^2}{f}+ r^2 d\Omega^2\,,\label{metric_Schwarzschild_psi}\\
\ti{g}_\mn dx^\mu dx^\nu
&=&- \fr{1}{1+\epsilon}\left(1-(1+ \epsilon) \frac{r_s}{r} \right) dt^2+ r^2d\Omega^2 \nonumber\\
&+& 2\sqrt{\frac{r_s}{r}} f^{-1} \frac{\epsilon dt dr }{1+\epsilon} + \left( 1 - \frac{r_s}{(1+\epsilon)r} \right) \frac{dr^2}{f^{2}} \,.
\label{metric_Schwarzschild}
\eeq
It should be noted that the disformal metric~$\tilde{g}_{\mu\nu}$ also describes a Schwarzschild spacetime but with a different horizon (up to a coordinate redefinition)~\cite{Dubovsky:2006vk,Mukohyama:2009rk} (see also Refs.~\cite{Takahashi:2019oxz,BenAchour:2020wiw}).
More concretely, the Schwarzschild horizon for the disformal metric is given by $\tilde{r}_s=(1+\epsilon)r_{\rm s}$, with $r_s$ being the Schwarzschild horizon for $g_{\mu\nu}$.\footnote{We can use the change of variables $dt \to \sqrt{1 + \epsilon}\,d\tilde{t} + \frac{\epsilon \sqrt{1-f}}{
f(-\epsilon + (1 + \epsilon) f)}dr$ with $f=1-r_s/r$, to bring the line element for $\ti{g}_\mn$ to the form
\be
\ti{g}_\mn dx^\mu dx^\nu=-\left(1-\frac{r_s(1+\epsilon)}{r}\right)d\tilde{t}^2+\left(1-\frac{r_s(1+\epsilon)}{r}\right)^{-1}dr^2+ r^2d\Omega^2\,,\nonumber
\ee
which is clearly a Schwarzschild spacetime.\label{footnote:metric}}
In our setup of $\epsilon>0$, the horizon for $\chi$ lies outside that for $\psi$.

In other words, the simple model above is a realization of a theory where different fields probe different geometries and different horizons. Although this is just one example among many possibilities, it provides a concrete theoretical framework in which one can explore the rich physics of a BH with nested multihorizons of the effective geometries. In the following, we will explore some of the dynamics of the theory.

\section{Numerical scheme\label{numerics}}

We note that a minimal coupling of the scalar fields to the curvature allows for the same vacuum solutions as general relativity. We will study the dynamics of $\psi$ and $\chi$ perturbatively in their amplitude, therefore fixing the background geometry to be the Schwarzschild one. The equations of motion for $\psi$ and $\chi$ are
\begin{align}
&\frac{1}{\sqrt{-g}} \partial_{\mu} (\sqrt{-g} g^{\mu \nu} \partial_{\nu} \psi) - \frac{9}{4} \frac{r_s}{r^3} (\epsilon_{12} \chi +\epsilon_{22} \psi) = 0\,,
\label{psi_eom}
\\
&\frac{1}{\sqrt{-\tilde{g}}} \partial_{\mu} (\sqrt{-\tilde{g}} \tilde{g}^{\mu \nu} \partial_{\nu} \chi) - \frac{9}{4} \frac{r_s}{r^3} (\epsilon_{11} \chi +\epsilon_{12} \psi) = 0\,.
\label{chi_eom}
\end{align}

The simplest possible solution to the above systems would be a nontrivial, static profile for the scalars. We have looked for regular, spherically symmetric solutions of $\psi,\chi$ down to the inner horizon at $r=r_s$, and we were unable to find any set of parameters for which this was possible. In other words, we find no linearized hairy BHs, except the trivial $\psi=\chi=0$ everywhere (and a special solution, which requires that $\psi,\chi$ are constants and only exist for a certain combination of coupling constants). We thus resort to a numerical evolution of time-dependent configurations.

\subsection{Setup}
\begin{figure}[t]
\centering
\includegraphics[width=0.5\linewidth]{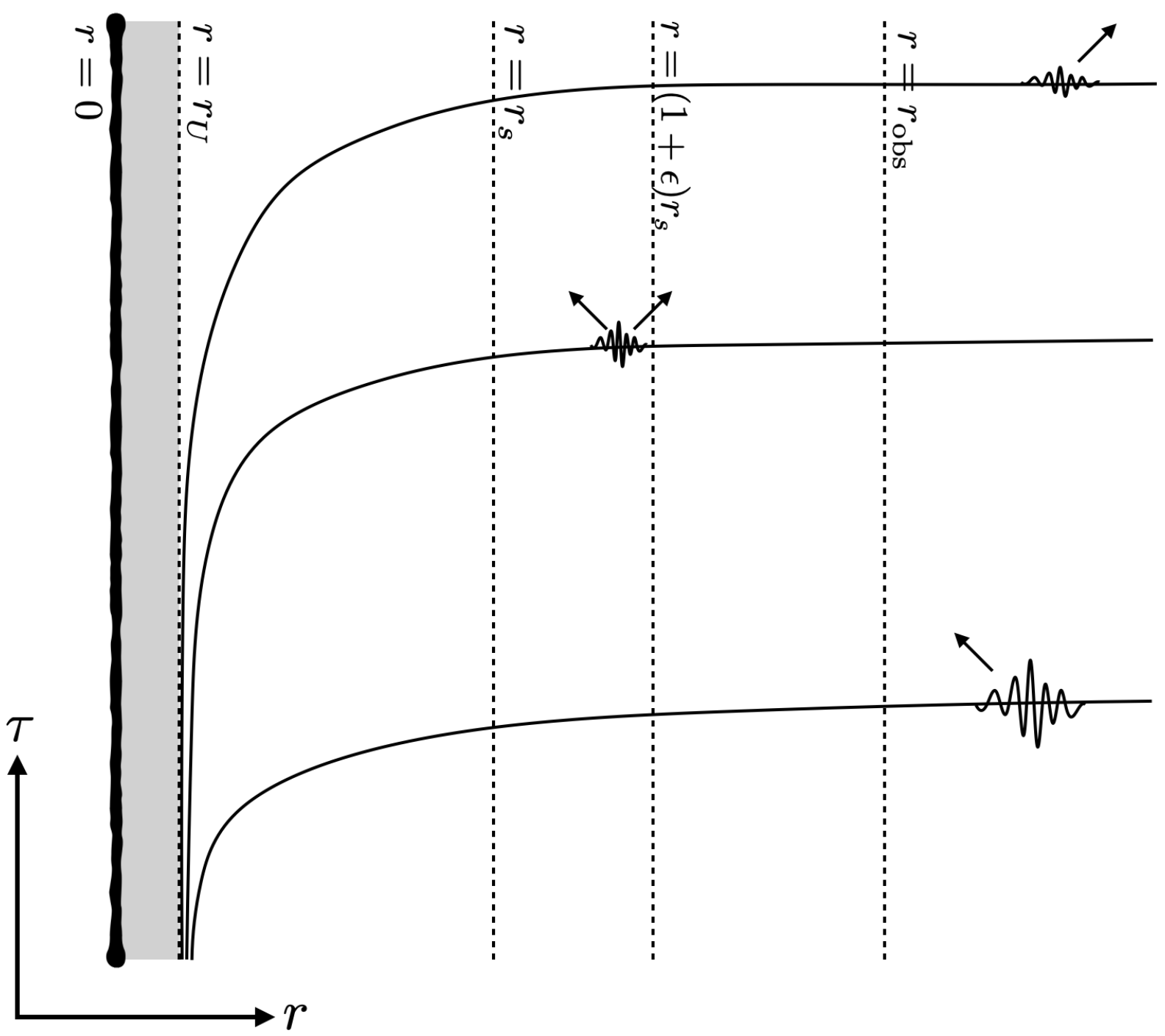}
\caption{Coordinates covering the interior of the two horizons and nicely avoid the singularity. Each curve corresponds to a constant-$T$ hypersurface.}
\label{pic_slices}
\end{figure}

Given the spherical symmetry of the background, we expand the scalar fields~$\psi$ and $\chi$ in their multipolar components,
\beq
\psi(\tau, \rho, \theta, \varphi)=\sum_{\ell m}\psi_{\ell m} (\tau, \rho) Y_{\ell m} (\theta, \varphi)\,,\qquad \chi(\tau, \rho, \theta, \varphi)=\sum_{\ell m}\chi_{\ell m} (\tau, \rho) Y_{\ell m} (\theta, \varphi)\,,
\eeq
where the $Y_{\ell m}$ are spherical harmonics on the 2-sphere. We will drop the $\ell m$ subscripts onward, with the understanding that we are always discussing the multipolar components of the field. Note also that the azimuthal number~$m$ never plays a role, due to the spherical symmetry of the background. We then numerically solve the two partial differential equations (PDEs)~\eqref{psi_eom} and \eqref{chi_eom}, which reduce to the following equations after performing the coordinate transformation introduced in Eq.~\eqref{tau_rho_coordinates}:
\begin{align}
&- \partial_{\tau}^2 \psi + \frac{r}{r_s} \partial_{\rho}^2 \psi - \frac{\ell (\ell+1)}{r^2} \psi+  \frac{1}{2r^2} \sqrt{\frac{r}{r_s}} \left[ 3 r_s \partial_{\tau} \psi + 5 r \partial_{\rho} \psi \right] - \frac{9}{4} \frac{r_s}{r^3} (\epsilon_{12} \chi +\epsilon_{22} \psi) = 0\,,\\
&- (1+\epsilon)\partial_{\tau}^2 \chi + \frac{r}{r_s} \partial_{\rho}^2 \chi - \frac{\ell (\ell+1)}{r^2} \chi+ \frac{1}{2r^2} \sqrt{\frac{r}{r_s}} \left[ 3 r_s (1+\epsilon) \partial_{\tau} \chi + 5 r \partial_{\rho} \chi \right] - \frac{9}{4} \frac{r_s}{r^3} (\epsilon_{11} \chi +\epsilon_{12} \psi) = 0\,.
\end{align}
Here, we have used the following relation to simplify the equations:
\begin{equation}
r = \left\{ \frac{3}{2} r_s^{1/2} (\rho -\tau) \right\}^{2/3}\,.
\end{equation}
The two fields~$\psi$ and $\chi$ have two distinct horizons.
We then solve the equations while covering the interior of both horizons. To avoid numerical instability that may arise near the singularity, we will use the following coordinates~$(T, r_{\ast})$ that nicely avoid it (see Fig.~\ref{pic_slices}),
\begin{align}
d\tau &= dT + \frac{1}{f} \left(U \sqrt{\frac{r_s}{r}} + V \right) dr_{\ast}\,, \label{tau2Trstar} \\
d\rho &= dT + \frac{1}{f} \left(U \sqrt{\frac{r}{r_s}} + V \right) dr_{\ast}\,, \label{rho2Trstar}
\end{align}
where
\begin{align}
U &\coloneqq 1-r_U/r\,,\\
V &\coloneqq -\sqrt{U^2 -f}\,,\\
r_{\ast} &\coloneqq r+r_U \log[(r-r_U)/r_s]\,,
\end{align}
and $r_U\,(\le r_s/2)$ is a constant controlling how close constant-$T$ slices can approach the singularity.\footnote{The relation between $(T,r_\ast)$ and $(t,r)$ is given by
    \begin{align}
    dt=dT+\frac{V}{f}dr_{\ast}\,, \qquad
    dr=Udr_{\ast}\,. \nonumber
    \end{align}
}
Note that the coefficients of $dr_{\ast}$ in \eqref{tau2Trstar} and \eqref{rho2Trstar} are regular at $r=r_s$.
We then have the following relations:
\begin{align}
\partial_{\rho} &= - \frac{r_s/r + \sqrt{r_s/r} (V/U)}{f} \partial_{T} + \frac{1}{U} \sqrt{\frac{r_s}{r}} \partial_{r_{\ast}}\,,\\
\partial_{\tau} &= \frac{1 + \sqrt{r_s/r} (V/U)}{f} \partial_{T} - \frac{1}{U} \sqrt{\frac{r_s}{r}} \partial_{r_{\ast}}\,.
\end{align}

In what follows, we will express all our results in units of $r_s$ and thus set $r_s=1$. With this, we find the equations of motion
\begin{align}
\begin{split}
&\left[ \partial_{T}^2 - f \partial_{r_{\ast}}^2 + 2V \partial_{T} \partial_{r_{\ast}} + U^2 \frac{\ell (\ell+1)}{r^2} \right] \psi + \frac{1}{U^2} \left[ fV' - \frac{UV}{2r^2} \right] \left( \partial_T +V \partial_{r_{\ast}} \right) \psi\\
&- \frac{2U}{r} \left[ 1-\frac{3}{4r} \right] \partial_{r_{\ast}} \psi +\frac{2UV}{r} \partial_T \psi + \frac{9}{4} \frac{U^2}{r^3} (\epsilon_{22} \psi + \epsilon_{12} \chi)= 0\,,
\label{eq:dynamics_1}
\end{split}\\
\begin{split}
&\left\{ \left( 1 + \epsilon W^2 \right) \partial_{T}^2 - f \partial_{r_{\ast}}^2 + 2V \partial_{T} \partial_{r_{\ast}} + U^2 \frac{\ell (\ell+1)}{r^2} \right\} \chi \\
&+ \frac{1}{U^2} \left[ fV' - \frac{UV}{2r^2} \right] \left( \partial_T +V \partial_{r_{\ast}} \right) \chi - \frac{2U}{r} \left[ 1-\frac{3}{4r} \right] \partial_{r_{\ast}} \chi +\frac{2UV}{r} \partial_T \chi + \frac{9}{4} \frac{U^2}{r^3} (\epsilon_{11} \chi + \epsilon_{12} \psi)\\
&+\epsilon \left[ -2 \sqrt{\frac{1}{r}} W \partial_T \partial_{r_{\ast}} - U \sqrt{\frac{1}{r}} \left( \frac{W}{U} \right)' \partial_T + \frac{1}{r} \partial_{r_{\ast}}^2 -\frac{2rU' +U^2}{2r^2 U} \partial_{r_{\ast}} \right.\\
&\left. - \frac{3U}{2r} \sqrt{\frac{1}{r}} \left( W \partial_T - \sqrt{\frac{1}{r}} \partial_{r_{\ast}} \right) \right] \chi = 0\,,\label{eq:dynamics_2}
\end{split}\\
W&\coloneqq \frac{U + \sqrt{1/r} V}{f}\,,
\end{align}
where a prime denotes differentiation with respect to $r_{\ast}$.
We note that the function~$W(r)$ is regular at $r=1$.
Indeed, we have
	\begin{align}
	W=1+\frac{r_U^2}{2(1-r_U)}+{\cal O}(r-1)\,.
	\end{align}
Setting $r_U=1/2$, the function $W(r)$ reduces to
\begin{equation}
W = \frac{2r + 2 \sqrt{r} +1}{2 \sqrt{r} (\sqrt{r} +1)}\,,
\label{Whalf}
\end{equation}
which is regular in the whole range of $r_{\ast}$. In the following, we take $r_U = 1/2$ and apply the formula~(\ref{Whalf}) to the wave equation.

The two second-order PDEs of (\ref{eq:dynamics_1}) and (\ref{eq:dynamics_2}) are decomposed into four first-order PDEs:
\begin{align}
\partial_T \psi &= \Pi_{\psi}\,,
\label{PDE1}\\
\begin{split}
\partial_T \Pi_{\psi} &= f \partial_{r_{\ast}}^2 \psi - 2V \partial_{r_{\ast}} \Pi_{\psi} - U^2 \frac{\ell (\ell+1)}{r^2} \psi - \frac{1}{U^2} \left[ fV' - \frac{UV}{2r^2} \right] \left( \Pi_{\psi} +V \partial_{r_{\ast}} \psi \right)\\
&+ \frac{2U}{r} \left[ 1-\frac{3}{4r} \right] \partial_{r_{\ast}} \psi -\frac{2UV}{r} \Pi_{\psi} - \frac{9}{4} \frac{U^2}{r^3} (\epsilon_{22} \psi + \epsilon_{12} \chi)= 0\,,
\end{split}
\label{PDE2}\\
\partial_T \chi &= \Pi_{\chi}\,,
\label{PDE3}\\
\begin{split}
(1+ \epsilon W^2) \partial_T \Pi_{\chi} &= f \partial_{r_{\ast}}^2 \chi - 2V \partial_{r_{\ast}} \Pi_{\chi} - U^2 \frac{\ell (\ell+1)}{r^2} \chi - \frac{1}{U^2} \left[ fV' - \frac{UV}{2r^2} \right] \left( \Pi_{\chi} +V \partial_{r_{\ast}} \chi \right)\\
&+ \frac{2U}{r} \left[ 1-\frac{3}{4r} \right] \partial_{r_{\ast}} \chi -\frac{2UV}{r} \Pi_{\chi} - \frac{9}{4} \frac{U^2}{r^3} (\epsilon_{11} \chi + \epsilon_{12} \psi) - \epsilon \left[ -2 W \sqrt{\frac{1}{r}} \partial_{r_{\ast}} \Pi_{\chi} + W \frac{U'}{U} \sqrt{\frac{1}{r}} \Pi_{\chi} \right. \\
&\left. -W' \sqrt{\frac{1}{r}} \Pi_{\chi} + \frac{1}{r} \partial_{r_{\ast}}^2 \chi - \frac{2rU'+U^2}{2r^2 U} \partial_{r_{\ast}} \chi - \frac{3 U}{2r} \sqrt{\frac{1}{r}} \left( W \Pi_{\chi} - \sqrt{\frac{1}{r}} \partial_{r_{\ast}} \chi \right)  \right] = 0\,.
\end{split}
\label{PDE4}
\end{align}
We numerically solve the PDEs of \eqref{PDE1}--\eqref{PDE4} with the following initial conditions,\footnote{The initial conditions may involve partially outgoing modes even when we set the wave packet at a distant region. This is not problematic in our computation of the reflectivity performed later as we read the time-domain data inside the initial position of the wave packets. That is, outgoing modes do not contaminate the time-domain data we use.}
\begin{align}
\psi(T=0, r_{\ast}) &= \psi_{\rm ini}(r_*)
\coloneqq A_{\psi} \cos (\Omega_{\psi} r_{\ast} + \delta_{\psi}) \exp \left[ - \frac{(r_{\ast} - r^{\psi}_{\ast, s})^2}{\sigma_{\psi}^2} \right],\\
\Pi_{\psi} (T=0, r_{\ast}) &= \partial_{r_*}\psi_{\rm ini}(r_*)
=- A_{\psi} \left( \Omega_{\psi} \sin (\Omega_{\psi} r_{\ast} + \delta_{\psi}) +2 \frac{(r_{\ast} - r^{\psi}_{\ast, s})}{\sigma_{\psi}^2} \cos (\Omega_{\psi} r_{\ast} + \delta_{\psi}) \right) \exp \left[ - \frac{(r_{\ast} - r^{\psi}_{\ast, s})^2}{\sigma_{\psi}^2} \right],\\
\chi(T=0, r_{\ast}) &= \chi_{\rm ini}(r_*)
\coloneqq A_{\chi} \cos (\Omega_{\chi} r_{\ast} + \delta_{\chi}) \exp \left[ - \frac{(r_{\ast} - r^{\chi}_{\ast, s})^2}{\sigma_{\chi}^2} \right],\\
\Pi_{\chi} (T=0, r_{\ast}) &= \partial_{r_*}\chi_{\rm ini}(r_*)
=- A_{\chi} \left( \Omega_{\chi} \sin (\Omega_{\chi} r_{\ast}  + \delta_{\chi}) +2 \frac{(r_{\ast} - r^{\chi}_{\ast, s})}{\sigma_{\chi}^2} \cos (\Omega_{\chi} r_{\ast}  + \delta_{\chi}) \right) \exp \left[ - \frac{(r_{\ast} - r^{\chi}_{\ast, s})^2}{\sigma_{\chi}^2} \right],
\end{align}
where $A_{\psi/ \chi}$, $\Omega_{\psi/\chi}$, $\sigma_{\psi/\chi}$, $\delta_{\psi/\chi}$, and $r^{\psi/\chi}_{\ast, s}$ are arbitrary parameters to specify the shape, initial position, and initial velocity of the input Gaussian wave packets. We control numerical high-frequency unstable modes by introducing the Kreiss-Oliger dissipation~\cite{kreiss1973methods}. We confirmed that this does not affect the benchmark tests we perform later. We also perform the resolution test which is described in the \hyperref[app_resolution]{Appendix}. 

\subsection{Diagnostics for energy extraction}
\label{sec_Diagnostics}
In this section, we explain how we discuss the energy extraction from a BH. For this purpose, we obtain the time-domain waveform of $\psi$ and $\chi$ at some fixed position~$r_{\ast}=r_{\ast,o}\gg 1$. We set the center of the initial Gaussian wave packet at $r_{\ast} = r^{\psi/\chi}_{\ast,s} > r_{\ast,o}$, so that we have an incoming wave packet in the early times, and after a while, we detect waves scattered by the double-horizon BH.

The energy flux~$T^{\tau \rho}$ evaluated at $r_{\ast} = r_{\ast, o}$ is given by
\begin{equation}
T^{\tau \rho} 
\sim - \left[ \partial_{T} \psi \partial_{r_{\ast}} \psi + (1+\epsilon) \partial_{T} \chi \partial_{r_{\ast}} \chi  \right],
\end{equation}
where we have approximated the expression by assuming $r_{\ast, o}\gg 1$ and have omitted an overall factor which is irrelevant in the following discussion.
Moreover, since $\psi \simeq \psi (T\mp r_{\ast})$ and $\chi \simeq \chi (T/\sqrt{1+\epsilon} \mp r_{\ast})$ at $r_{\ast} = r_{\ast, o} \gg 1$, the expression for $T^{\tau\rho}$ can be rewritten as
\begin{align}
T^{\tau\rho}\sim \pm \left[ (\partial_{T} \psi)^2 + (1+\epsilon)^{3/2} (\partial_{T} \chi)^2  \right],
\end{align}
with the plus (minus) sign in front corresponding to the outgoing (incoming) flux.
At a distant region, the total energy flux is given by the sum of the two energy fluxes for $\psi$ and $\chi$. To compute the energy flux of incoming and scattered (outgoing) flux, we separate the energy flux into incoming and outgoing modes in the time domain by introducing the truncation parameters $T = T_{\rm cut1/cut 2}$ as
\begin{align}
&\tilde{\psi}_{\rm in} \coloneqq \int^{T_{\rm cut1}}_0 dTe^{i\omega T} (\partial_T \psi)\,, \qquad
\tilde{\psi}_{\rm out} \coloneqq \int^{T_{\rm end}}_{T_{\rm cut1}} dTe^{i\omega T} (\partial_T \psi)\,,\\
&\tilde{\chi}_{\rm in} \coloneqq \int_{0}^{T_{\rm cut2}} dTe^{i\omega T} (\partial_T \chi)\,, \qquad
\tilde{\chi}_{\rm out} \coloneqq \int_{T_{\rm cut2}}^{T_{\rm end}} dTe^{i\omega T} (\partial_T \chi)\,,
\end{align}
to obtain the spectral amplitude of each signal. Then, the energy flux per frequency is
\begin{align}
&\frac{dE_{\rm in}}{d\omega} \sim  \left[ |\tilde{\psi}_{\rm in}|^2 + (1+\epsilon)^{3/2} |\tilde{\chi}_{\rm in}|^2 \right],\\
&\frac{dE_{\rm out}}{d\omega} \sim  \left[ |\tilde{\psi}_{\rm out}|^2 + (1+\epsilon)^{3/2} |\tilde{\chi}_{\rm out}|^2 \right].
\end{align}
Finally, we compute the net reflectivity~${\cal R}^2$ as
\begin{equation}
{\cal R}^2 \coloneqq {\cal R}_{\psi}^2 +  {\cal R}_{\chi}^2 = \frac{|\tilde{\psi}_{\rm out}|^2 + (1+\epsilon)^{3/2} |\tilde{\chi}_{\rm out}|^2}{|\tilde{\psi}_{\rm in}|^2 + (1+\epsilon)^{3/2} |\tilde{\chi}_{\rm in}|^2}\,,
\label{reflectivity_dif}
\end{equation}
where
\begin{equation}
{\cal R}_{\psi}^2 \coloneqq \frac{|\tilde{\psi}_{\rm out}|^2}{|\tilde{\psi}_{\rm in}|^2 + (1+\epsilon)^{3/2} |\tilde{\chi}_{\rm in}|^2}\,, \qquad
{\cal R}_{\chi}^2 \coloneqq \frac{ (1+\epsilon)^{3/2} |\tilde{\chi}_{\rm out}|^2 }{|\tilde{\psi}_{\rm in}|^2 + (1+\epsilon)^{3/2} |\tilde{\chi}_{\rm in}|^2}\,.
\end{equation}
When the net reflectivity exceeds unity for some range of $\omega$, we conclude that the energy extraction from a BH occurs. We shall apply the above methodology to our model in Sec.~\ref{sec_SR}.

\section{Results\label{sec:results}}

\subsection{Benchmarks}
\label{sec_benchmark}
%
\begin{figure}[b]
\centering
\includegraphics[width=0.9\linewidth]{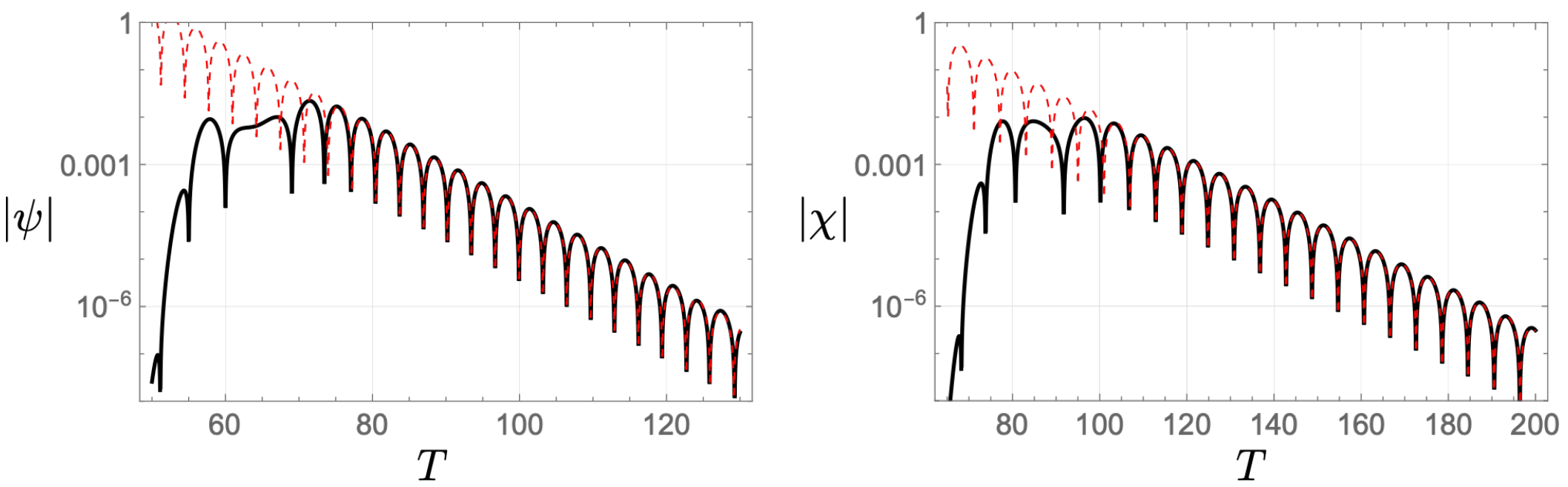}
\caption{Evolution of a Gaussian wave packet of the fields~$\psi$ and $\chi$, governed by the dynamical equations~\eqref{eq:dynamics_1} and \eqref{eq:dynamics_2}, is shown in the log-linear scale. We set $\epsilon_{ij}=0$ and $\epsilon = 0.5$. The ringdown of the field~$\psi$ (black solid) is consistent with the fundamental QNM frequency~\eqref{fundamental_qnm_zero_couplings} (red dashed).
The field~$\chi$, on the other hand, couples to an effective geometry with horizon radius~$r_s(1+\epsilon)$ and thus decays slower. We show in the right panel that its decay is equivalent to that of $\psi$ with the appropriate rescalings, $T \to T/\sqrt{1+\epsilon}$ and $\omega_{\ell mn} \to \omega_{\ell mn}/(1+\epsilon)$. Throughout the discussion, we set units such that $r_s=1$.
}
\label{pic_test1}
\end{figure}
\begin{figure}[h]
\centering
\includegraphics[width=0.5\linewidth]{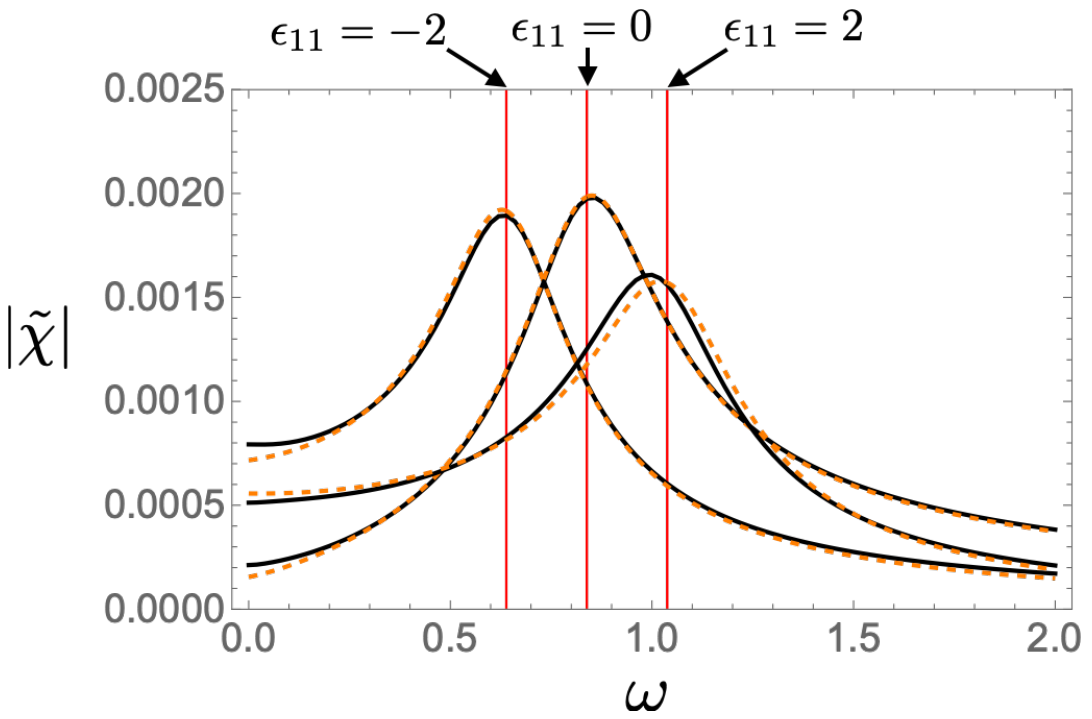}
\caption{Energy spectrum for the field $\chi$, for $\epsilon=0.1$, together with a least-squares fit analysis. Red solid lines indicate the value of $\text{Re} (\tilde{\omega}_{\rm Q})$ for $\epsilon_{11}=\pm 2$ and $0$. 
We find excellent agreement with the modified QNM~$\omega_{\rm Q}$ derived in (\ref{omega-q}).}
\label{pic_massive_qnm}
\end{figure}
We start with a few benchmark tests on the equations of motion and our numerical implementation. Throughout the paper, unless otherwise mentioned, we consider the quadrupolar mode ($\ell =2$), $r_{\ast, o} = 60$, and set the initial Gaussian wave packet with $\sigma_{\psi} = 2 = \sigma_{\chi}$, $\Omega_{\psi} = 1 =\Omega_{\chi}$, and $\delta_{\psi} = 0 =\delta_{\chi}$. Also, we set $A_{\psi} = 1 = A_{\chi}$ in Secs.~\ref{sec_benchmark} and \ref{sec_QNM}. To model a situation where the fields are simultaneously excited near the horizons, we set its position at $r^{\psi/\chi}_{\ast,s} = 5$ unless otherwise noted. Throughout this section, we turn off the coupling between the fields by setting $\epsilon_{12}=0$.

\subsubsection{Ringdown}
In Schwarzschild-like coordinates, with the coupling turned off ($\epsilon_{12}=0$) and with $r\psi(t,r)=e^{-i\omega t}\psi_a(r)$, $r\chi(t,r)=e^{-i\omega t}\psi_b(r)$, we find
\beq
&&f^2\psi''_a+ff'\psi'_a+\left(\omega^2-V_a\right)\psi_a=0\,,\\
&&V_a=f\left(\frac{\ell(\ell+1)}{r^2}+\frac{r_s}{r^3}(1+9\epsilon_{22}/4)\right),\label{eq:psi_no_coupling}
\eeq
and a similar (but more complicated) equation for $\psi_b$. This equation has the same form as that of massless fields, and standard methods can be used to calculate QNMs, for example, a continued-fractions approach~\cite{Leaver:1985ax,Berti:2009kk}.\footnote{In particular, we find the same continued fraction representation as Leaver~\cite{Leaver:1985ax}, with his $\epsilon\to -(1+9\epsilon_{22}/4)$.} In particular, for $\epsilon_{22}=0$, one finds a classical result,
\be
r_s\omega_{220}=0.967288 - 0.193518i\,,\label{fundamental_qnm_zero_couplings}
\ee
for the fundamental QNM frequency of a quadrupolar ($\ell=2$) mode, where a QNM frequency, $\omega= \omega_{\ell mn}$, is labeled by angular modes $(\ell,m)$ and the overtone number $n$.\footnote{As mentioned earlier, the QNMs are independent of the azimuthal number~$m$, and we set $m=2$ just for concreteness.}
For small $\epsilon_{22}$, we find with a continued-fraction approach,
\be
r_s(\omega_{\rm Q}-\omega_{220})=\frac{9\epsilon_{22}}{4}(0.05112-0.0032973i)\,,
\label{omega-q}
\ee
which agrees with the parametrized results of Ref.~\cite{Cardoso:2019mqo} (in their notation, $\beta_3^s=9\epsilon_{22}/4$). 

Our time-domain results are summarized in Fig.~\ref{pic_test1}. We scatter a Gaussian wave packet in the spacetime, as described previously. After an immediate response as a consequence of direct on-light-cone propagation, the field configuration relaxes in a series of exponentially damped sinusoids. This stage is known as ringdown stage and is dominated by the BH QNMs~\cite{Berti:2009kk}.
Figure~\ref{pic_test1} shows a clear ringdown waveform for both fields. Fitting our numerical data, we recover the prediction~\eqref{fundamental_qnm_zero_couplings} with very good accuracy when $\epsilon_{22}=0$ for the $\psi$ field.

The above result concerns the field $\psi$. It is also apparent that the field~$\chi$ decays slower. In fact, as we pointed out before in footnote~\ref{footnote:metric}, the effective metric probed by the field~$\chi$ is simply a rescaled version of the Schwarzschild metric. In particular, $\chi$ is coupling to a geometry with a Schwarzschild radius~$r_s(1+\epsilon)$ and time coordinate rescaled by $\sqrt{1+\epsilon}$. Taking the rescalings involved, we find that the mode frequency 
$\tilde{\omega}$ of $\chi$ should be related to that of $\psi$ via 
\be
\tilde{\omega}=\frac{\omega}{(1+\epsilon)^{3/2}}\,.\label{w_chi}
\ee
The right panel of Fig.~\ref{pic_test1} compares the decay of $\chi$ against the prediction~\eqref{w_chi}.
The agreement is excellent and further validated in Fig.~\ref{pic_massive_qnm}, where we extend the comparison to other values of $\epsilon_{11}$ and do it in the frequency domain directly. We introduce the Lorentzian functions,
\begin{equation}
\frac{A}{\omega - \tilde{\omega}_{\rm Q}} + \frac{B}{\omega +\tilde{\omega}^{\ast}_{\rm Q}}\,,
\label{lorentzian}
\end{equation}
where $\tilde{\omega}_{\rm Q} \coloneqq \omega_{\rm Q}/(1+\epsilon)^{3/2}$. We then fit the model~(\ref{lorentzian}) with the Fourier transform of $\chi$, denoted as $\tilde{\chi}$ in Fig.~\ref{pic_massive_qnm}, by using the least-square method to determine the fitting parameters~$A$ and $B$. The two free parameters are real and are relevant to the amplitude and phase. The agreement is excellent as the peak (real part of QNM frequency) and broadness (relevant to the quality factor) of $\tilde{\chi}$ fit very well with the Lorentzian function with $\tilde{\omega}_{\rm Q}$.

\subsubsection{Tidal numbers}

It is also straightforward to calculate the tidal response in the decoupling limit. In the standard massless scalar in a Schwarzschild background, the tidal Love numbers are zero. By contrast, in our setup they are nontrivial. For a quadrupolar field, for example, we find a ``running'' of the coupling, with the regular solution at the horizon
\begin{align}
\psi_a=r^3\left[1+{\cal O}\left(r^{-1}\right)\right]-\frac{1}{180}\cdot\frac{9\epsilon_{22}}{4}r^{-2}\left[\log r+{\cal O}(1)\right]\left[1+{\cal O}\left(r^{-1}\right)\right]\,,
\end{align}
a result which can also be read off from the parametrized study of Ref.~\cite{Katagiri:2023umb}.

\subsubsection{Instability}
%
\begin{figure}[t]
\centering
\includegraphics[width=0.5\linewidth]{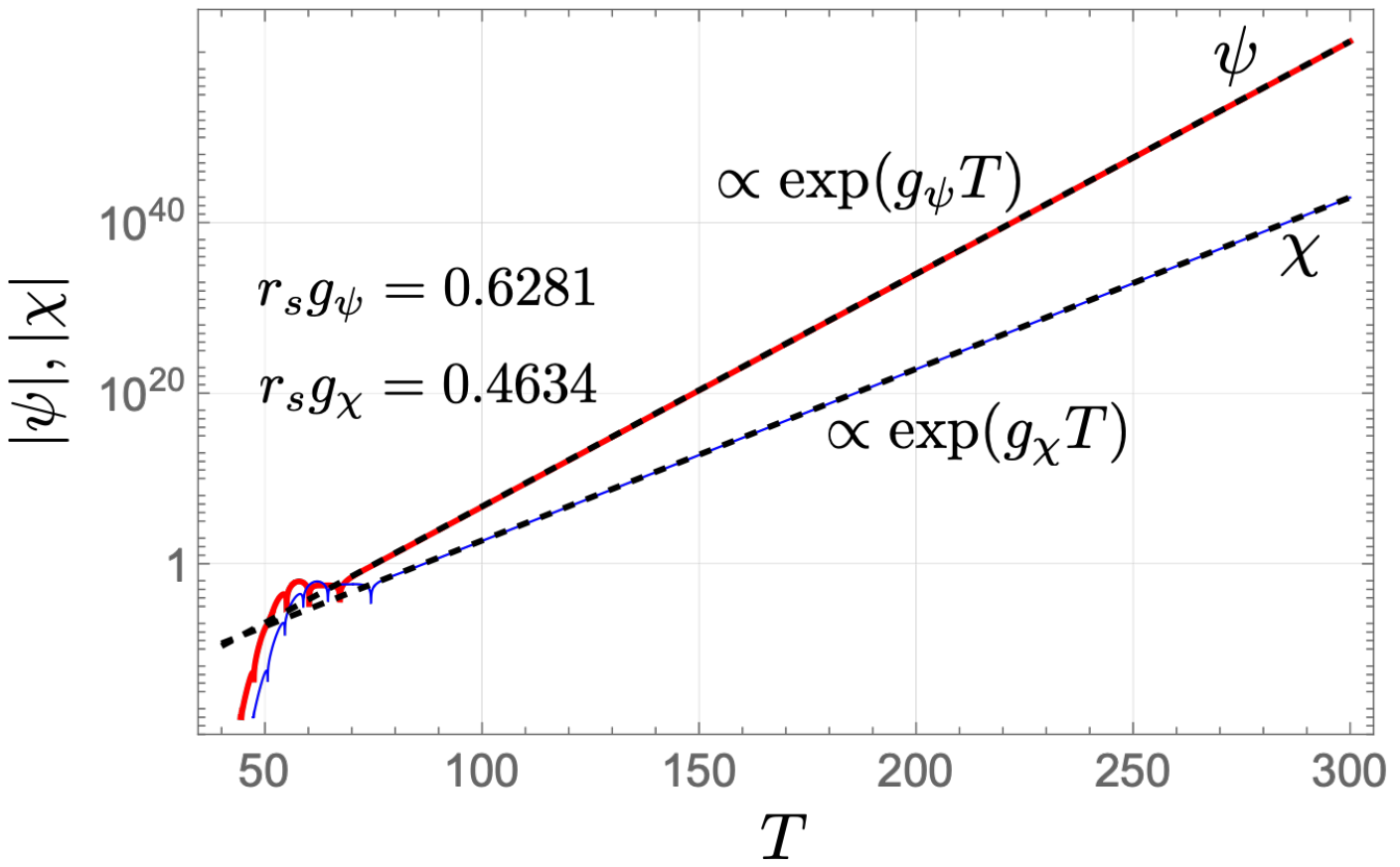}
\caption{Evolution of a Gaussian wave packet of the fields~$\psi$ and $\chi$, governed by the dynamical equations~\eqref{eq:dynamics_1} and \eqref{eq:dynamics_2}, is shown in the log-linear scale. We set $\epsilon= 0.1$ and $\epsilon_{11} = \epsilon_{22} =-7$.
Both $\epsilon_{11}$ and $\epsilon_{22}$ are masslike parameters, and large negative values trigger a tachyoniclike instability.
The fields grow exponentially at late times with a rate $r_s g_{\psi}=0.6281$ and $r_s g_{\chi}=0.4634$ as shown in the figures. Those values are read by the least-square method. The instability rate of $\psi$ is in excellent agreement with $r_s g_{\psi}=0.6283$ and $r_s g_{\chi}=0.4637$, obtained from a continued-fraction solution for the unstable mode of the corresponding equation in the frequency domain, Eq.~\eqref{unstable_tachyon}.
}
\label{pic_test2}
\end{figure}
The results so far are perturbative in $\epsilon_{22}$ (or $\epsilon_{11}$). For large negative values of this constant, we find an unstable mode of Eq.~\eqref{eq:psi_no_coupling}. In fact, the effective potential~$V_a$ becomes negative in some region of $r$ and a sufficient (but not necessary) condition for an unstable mode of $\psi$ to appear is that~\cite{Buell:1995,Cardoso:2009cnd}
\be
\int_{r_s}^\infty dr\frac{V_a}{f}<0\,,
\ee
which amounts to the condition that $2\ell(\ell+1)+1+9\epsilon_{22}/4<0$. For $\ell=2$ for example, it is {\it sufficient} that $\epsilon_{22}<-5.8$ for an unstable mode to appear. A similar analysis reveals that a sufficient condition for an unstable mode of $\chi$ to appear is given by $\epsilon_{11}/(1+\epsilon)<-5.8$.

A continued-fraction solution yields accurate values for the unstable mode, which corresponds to a purely imaginary value of $\omega$ with a positive imaginary part. We define the growth rate for $\psi$ and $\chi$ as $g_{\psi/\chi}\coloneqq -i\omega$. For $\epsilon=0.1$, $\ell=2$, and
$\epsilon_{22}=-7=\epsilon_{11}$, we find 
\be
r_sg_{\psi} = 0.6283\,, \qquad
r_sg_{\chi} = 0.4637\,.
\label{unstable_tachyon}
\ee
Our time-domain results are shown in Fig.~\ref{pic_test2}, which shows a clear exponential growth of both fields. The growth rate is in very good agreement with the prediction~\eqref{unstable_tachyon}.
\subsubsection{Graybody factors}
\begin{figure}[h]
\centering
\includegraphics[width=0.5\linewidth]{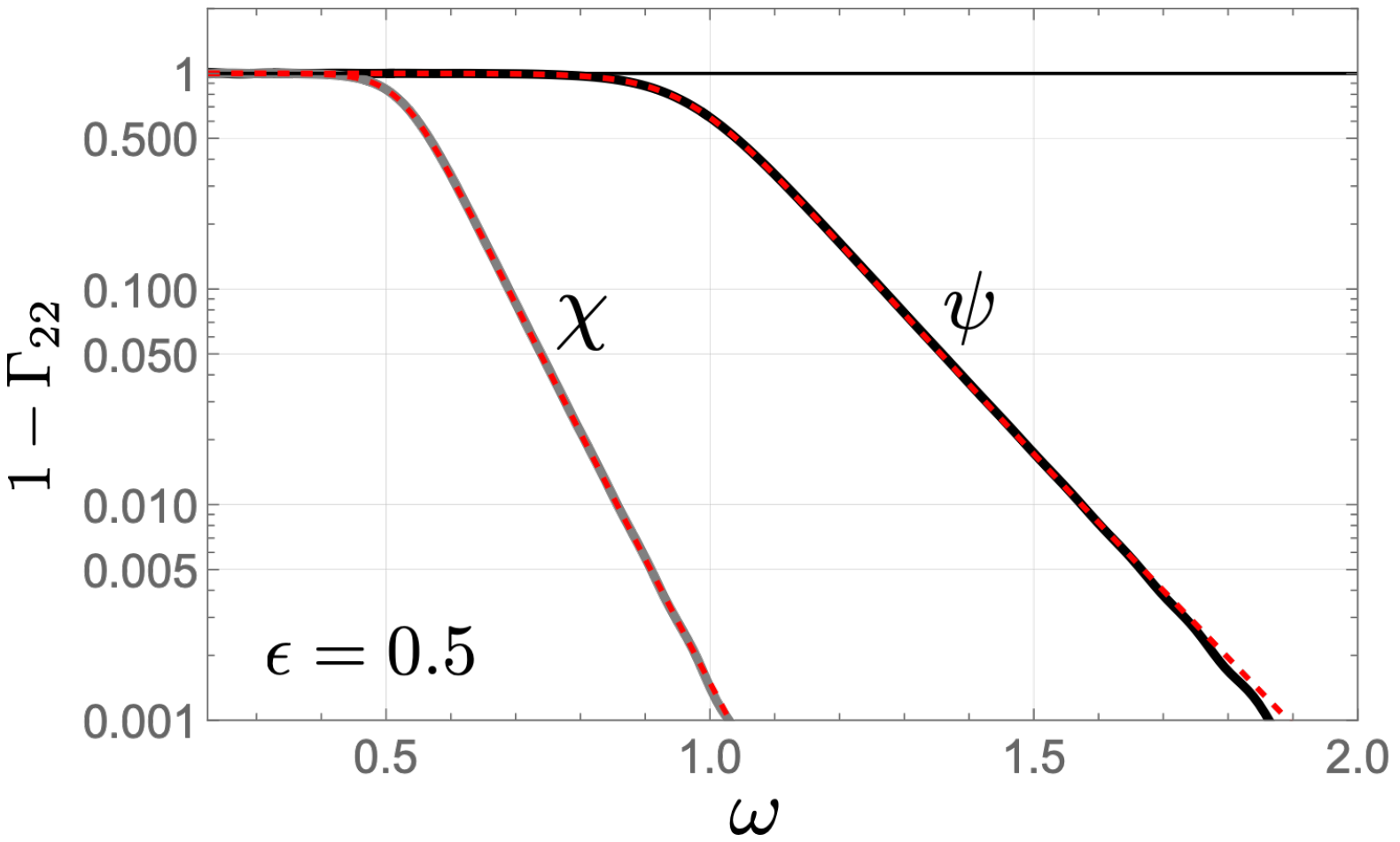}
\caption{The graybody factors for $\psi$ (black solid) and $\chi$ (gray solid) are shown in the log-linear scale. We set $\epsilon_{11}=\epsilon_{22}=\epsilon_{12}=0$ and $\ell=2$. The graybody factors obtained by the Heun function technique are shown with red dashed lines. We set $T_{\rm cut1}=82$ and $T_{\rm cut2}=99$.
}
\label{pic_grey}
\end{figure}
A measure of the permeability of the angular momentum barrier to incoming waves is the transmission coefficients in a scattering experiment. 
Given regular fields at the horizon~$r=r_s$, a scattering experiment for field~$\psi$, say, consists of imposing the asymptotic behavior
\beq
\tilde{\psi}=\left\{
\begin{array}{l}
A_{\rm out}(\omega)e^{i\omega x}+A_{\rm in}(\omega)e^{-i\omega x}\,,\quad x\to \infty\,,\\
A_{\rm T}e^{-i\omega x}\,,\qquad x\to -\infty\,.
\end{array}
\right.
\eeq
Here, $x$ is the tortoise coordinate defined by $dr/dx=1-r_s/r$. A similar expression holds for $\tilde{\chi}$, the Fourier transform of $\chi$. Each of the amplitudes~$A_{\rm out}, A_{\rm in}, A_{\rm T}$ is arbitrary, since the problem is linear, but their ratio is fixed by boundary condition at the horizon. As such, one defines the BH graybody factor~$\Gamma_{\ell m}$ as 
\beq
\Gamma_{\ell m} \coloneqq 1-|A_{\rm out}(\omega)/A_{\rm in}(\omega)|^2\,.
\eeq
This is an important quantity that characterizes classical BHs and also their quantum spectrum, in particular for the emission rate of Hawking radiation~\cite{Page:1976df}.
We computed the graybody factors for $\psi$ and $\chi$ with $\epsilon_{11}=\epsilon_{22}=\epsilon_{12}=0$ and $\epsilon=0.5$. We compute the energy flux of the injected Gaussian wave packet and that of the reflected waves by following the methodology presented in Sec.~\ref{numerics}. To obtain the injected and reflected waveforms in the time domain, we set $r^{\psi/\chi}_{\ast,s} = 80$ here for both $\psi$ and $\chi$ and read the amplitudes at $r_{\ast,o} = 60$. Our results shown in Fig.~\ref{pic_grey} are in excellent agreement with graybody factors computed with a different analytic technique, using the Heun function~\cite{Gregory:2021ozs}.

\subsection{QNM excitation in ringdown}
\label{sec_QNM}
%
\begin{figure}[h]
\centering
\includegraphics[width=0.95\linewidth]{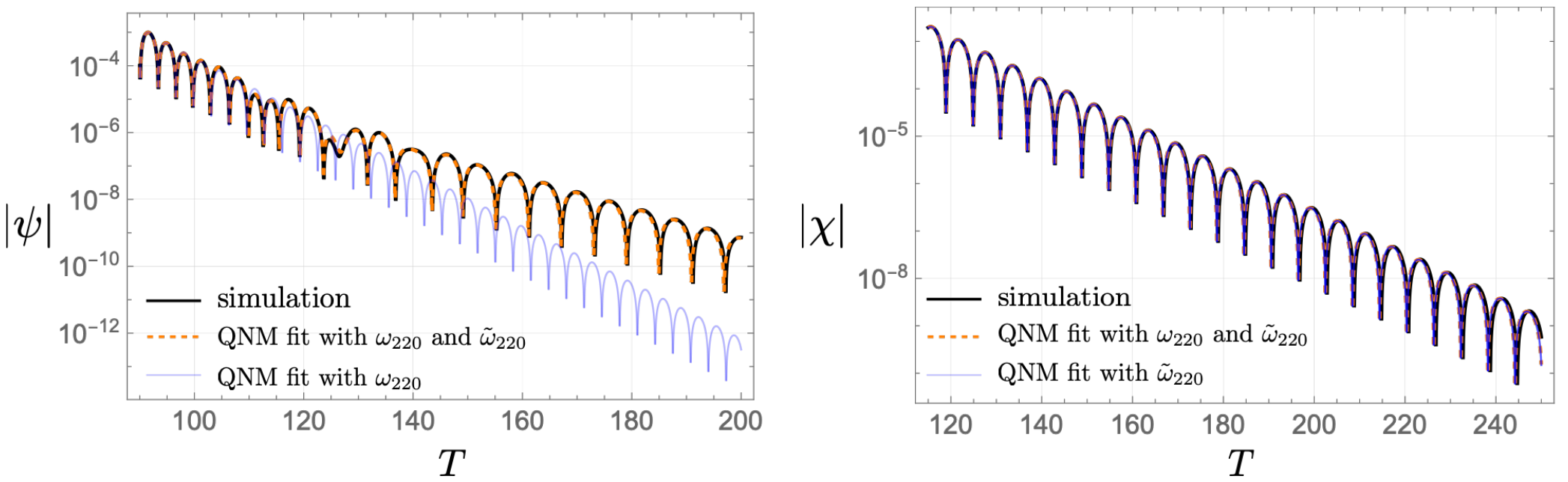}
\caption{Evolution of a Gaussian wave packet in a geometry with $\epsilon_{11}=\epsilon_{22}=0$, $\epsilon_{12}=0.5$, and $\epsilon = 0.5$. We show the evolution of $\psi$ on the left panel and of $\chi$ on the right panel in the log-linear scale. The least-square fit is performed with the data in $T \geq 90$ and in $T \geq 115$ for $\psi$ and $\chi$, respectively. We also show the QNM models with a sum of the two fundamental modes~$\omega_{220}$ and $\tilde{\omega}_{220}$ (orange dashed) and with $\omega_{220}$ only (blue thin solid).} 
\label{pic_ringdown_superposed}
\end{figure}
\begin{figure}[h]
\centering
\includegraphics[width=0.9\linewidth]{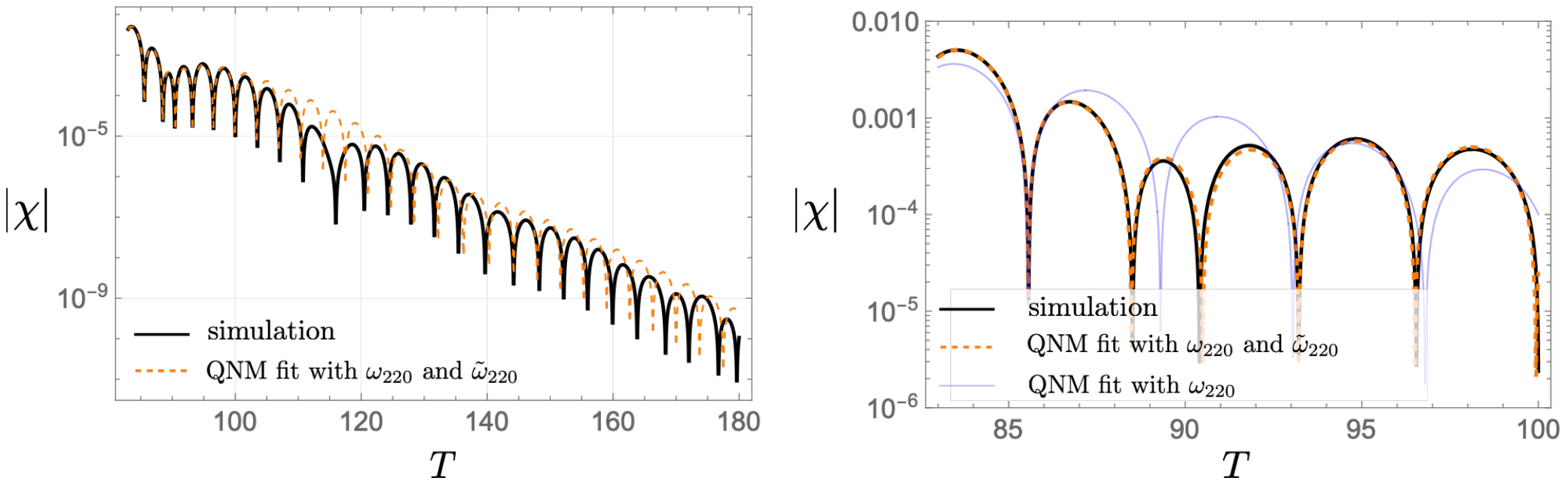}
\caption{Waveform of $\chi$ from the scattering of a wave packet in a spacetime with $\epsilon=0.1$, $\epsilon_{12}=1$, and $\epsilon_{11}=\epsilon_{22}=0$ (black solid line). The two light rings are located outside the external horizon of $\chi$, and therefore both fields are modulated by QNMs of each other. We fit with two fundamental modes for the observable and hidden sectors, $\omega_{220}$ and $\tilde{\omega}_{220}$ [see Eq.~\eqref{w_chi}], by using the least-square fit (orange dashed). The modulation at early times is well modeled by sum of the two fundamental modes. The right panel is an enlarged version of the left one. Using only one fundamental mode does not provide a good match (blue thin solid line). Both panels are shown in the log-linear scale.}
\label{pic_ringdown_interior_info}
\end{figure}
\begin{figure}[h]
\centering
\includegraphics[width=0.9\linewidth]{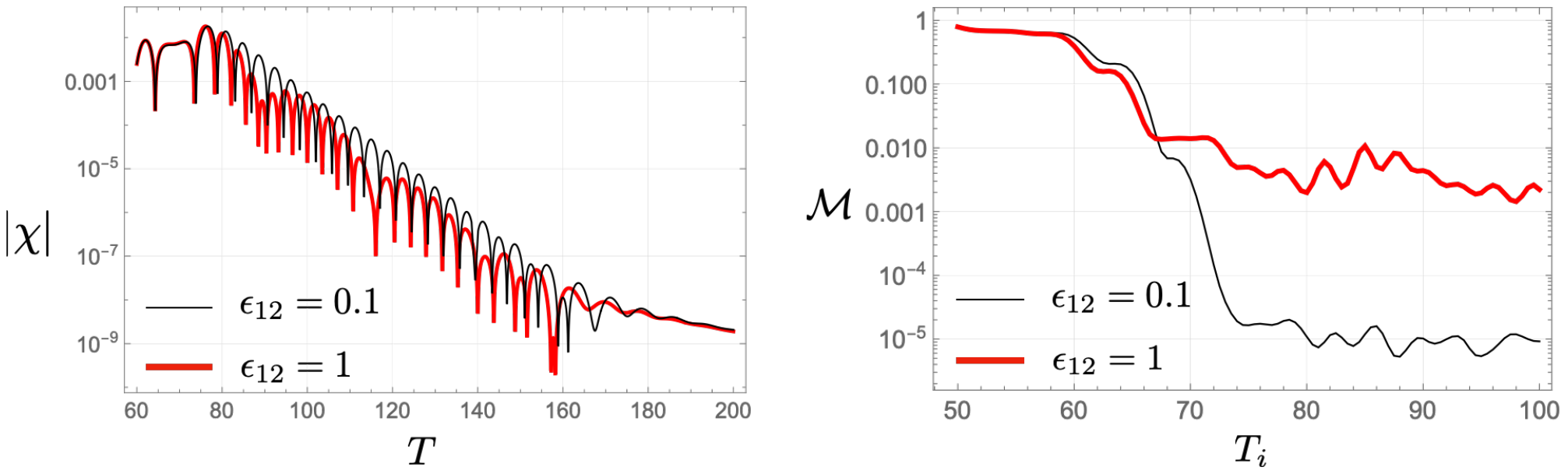}
\caption{{\bf Left panel:} waveforms for the observable sector $\chi$, for couplings $\epsilon_{12}=0.1, 1$ (black thin and red thick lines, respectively). We set $\epsilon_{11}=0=\epsilon_{22}$ and $\epsilon = 0.1$ for both cases. {\bf Right panel:} the mismatch between the ringdown waveform and our QNM model consisting of four $\chi$ QNMs up to $n=3$. We perform the least-square fit with the different start times of fit, $T_i$. Both panels are shown in the log-linear scale.}
\label{pic_ringdown_interior_info2}
\end{figure}
\begin{figure}[t]
\centering
\includegraphics[width=0.9\linewidth]{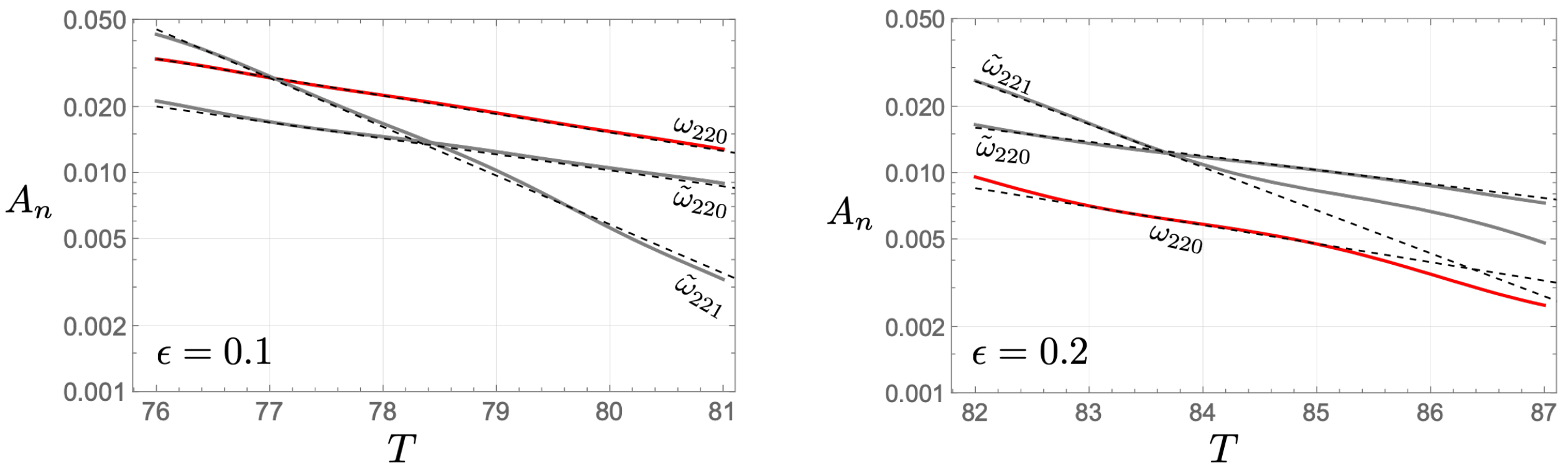}
\caption{QNM amplitudes read by the least-square fit with the three QNMs: the hidden sector's fundamental mode (red), $\omega_{220}$, and the observable sector's fundamental mode and first tone (gray), $\tilde{\omega}_{220}$ and $\tilde{\omega}_{221}$. The value of $\epsilon$ is set to $\epsilon = 0.1$ (left) and to $\epsilon = 0.2$ (right). The start time of fit is $T=76$ (left) and $T=82$ (right). We set $\epsilon_{11}=\epsilon_{22}=0$ and $\epsilon_{12}=1$. Both panels are shown in the log-linear scale.}
\label{pic_QNM_amps}
\end{figure}

In the benchmark tests performed in the previous section, we neglected the coupling between the fields~$\psi$ and $\chi$. This is a fundamental piece of our setup, and we now discuss its impact on the dynamics of linearized fields. It is important at this stage to reiterate that both fields propagate on a Schwarzschild background, one with radius~$r_s$ and the other with $r_s(1+\epsilon)$. It is also important to highlight that, although we focus on two coupled scalars, ultimately we want to draw conclusions for gravity as well. As such, observable gravitational waves can be associated either with $g_{\mu \nu}$ or $\tilde{g}_{\mu \nu}$, and we may assume that either $\psi$ or $\chi$ is in an observable sector, while the other is in a hidden sector, with negligible couplings to the Standard Model of particle physics. As we now show, there are exciting imprints of the hidden sector in the observable sector, specifically in the relaxation stage. Our results of scattering Gaussian wave packets in this setup are summarized in Figs.~\ref{pic_ringdown_superposed}--\ref{pic_QNM_amps}.

Figure~\ref{pic_ringdown_superposed} shows the late-time relaxation of field~$\psi$ (left) and of field~$\chi$ (right), for $\epsilon_{11}=\epsilon_{22}=0$, $\epsilon_{12}=0.5$, and $\epsilon = 0.5$. Notice that the fields are coupled now, via nonvanishing $\epsilon_{12}$. After the direct signal (not shown), field~$\psi$ decays in two different stages, apparent from the figure. The first stage corresponds to interaction with the light ring\footnote{Because of the coupling between $\psi$ and $\chi$, the potential in the perturbation equations takes a matrix form. Nevertheless, when the coupling is weak, we use the terminology ``light ring'' to refer to the position of the angular momentum barrier without coupling corrections. At the leading order, the ringdown can be approximately modeled by a superposition of the QNMs in $g_{\mu \nu}$ and those in $\tilde{g}_{\mu \nu}$, e.g., as is shown in Fig.~\ref{pic_ringdown_superposed}.} in ``its own'' spacetime, i.e., in geometry~$g_{\mu\nu}$ of Eq.~\eqref{metric_Schwarzschild_psi}, at $r=3r_s/2$. This explains the first exponentially damped stage. However, the coupling to $\chi$ implies that the field~$\psi$ also has access to the exterior light ring, that of geometry~(\ref{metric_Schwarzschild}). In other words, the field~$\psi$ is sensitive to the ringdown of field~$\chi$. 
The effective geometry of $\chi$ corresponds to a larger-mass BH, and hence rings at lower frequencies, as we showed already. Indeed, we show in Fig.~\ref{pic_ringdown_superposed} also the result of fitting the signal with the two fundamental modes of $\omega_{220}$ and $\tilde{\omega}_{220} \coloneqq \omega_{220}/(1+\epsilon)^{3/2}$ in Eq.~\eqref{w_chi}. The agreement is excellent.

On the other hand, the field~$\chi$ has no information on the inner light ring
(to leading order at least). In fact, for this choice of parameters, the light ring for $\psi$ overlaps with the horizon for $\chi$. This explains why $\chi$ relaxes as a clean damped sinusoid corresponding to the fundamental quadrupolar mode of the outer horizon~\eqref{w_chi}. Such a pure ringdown is apparent in the right panel of Fig.~\ref{pic_ringdown_superposed}.

For $\epsilon<0.5$ on the other hand, the light ring associated to $\psi$ lies on the {\it exterior} of the $\chi$ horizon, and one would expect both fields to carry imprints of the two light rings. Indeed, this feature is apparent in Fig.~\ref{pic_ringdown_interior_info}, where we evolve both fields but now for $\epsilon=0.1$. The two fundamental modes of $\psi$ and $\chi$ are excited in the ringdown of $\chi$ and lead to the modulation pattern as shown in Fig.~\ref{pic_ringdown_interior_info}.

What we have shown is that, given an ``observable field''~$\chi$ (i.e., one interacting with our detectors), its relaxation properties can show imprints of inner horizons for invisible fields~$\psi$ with which it couples. To quantify the effect, we compute the mismatch~${\cal M}$ between the ringdown waveform and the 
superposition of $\chi$ QNMs.
The mismatch~${\cal M}$ is defined as
\begin{align}
{\cal M} &\coloneqq 1- \left| \frac{\braket{\chi_{\rm Q}|\chi}}{\sqrt{\braket{\chi_{\rm Q}|\chi_{\rm Q}} \braket{\chi|\chi}}} \right|,\\
\chi_{\rm Q} &\coloneqq \sum_{n=0}^{n_{\rm max}} A_{n} \cos{(\tilde{\omega}_{n} T + \delta_{n})}\,,
\end{align}
and we use up to the third overtone, i.e., $n_{\rm max}=3$. The inner product between two functions $a(T)$ and $b(T)$ is given by
\begin{equation}
\braket{a|b} \coloneqq \int_{T_i}^{T_f} dT' a(T') b(T')\,,
\end{equation}
with $T_f = 200$ at which the amplitude of $\chi$ is well suppressed.
The fitting parameters~$A_{n}$ and $\delta_{n}$ are obtained by using the least-square method.
Results are shown in Fig.~\ref{pic_ringdown_interior_info2}.
We find that, for $\epsilon_{12}=0.1$, there is no modulation in ringdown and the least value of the mismatch is of the order of ${\cal M} \sim {\cal O}(10^{-5})$. On the other hand, a larger interaction with $\epsilon_{12}=1$ leads to the modulation in ringdown and the mismatch is ${\cal M} \sim {\cal O}(10^{-2})$. It means that the hidden sector~$\psi$ can be probed from the ringdown of an observable sector and may affect the QNM measurement utilizing the fitting analysis.

We also read the amplitude of each QNM. Concerning a possible overfitting, we use the three QNMs here: the fundamental mode and the first overtone for $\chi$, $\tilde{\omega}_{220}$ and $\tilde{\omega}_{221}$, and another fundamental mode for $\psi$, $\omega_{220}$. 
We then find that the fundamental mode for the hidden sector can be dominant as is shown in the left panel of Fig.~\ref{pic_QNM_amps}. Indeed, when the two horizons are close to each other, say, $\epsilon=0.1$, the amplitude for the hidden sector's fundamental mode is larger than the QNM excitation of the observable sector~$\chi$. For a larger separation between the horizons, say, $\epsilon=0.2$, the fundamental mode for the hidden sector is less dominant, as is shown in the right panel of Fig.~\ref{pic_QNM_amps}. The QNMs of $\psi$ and $\chi$ are more or less affected by the interaction controlled by $\epsilon_{12}$. Nevertheless, this result shows that our fits extract the amplitudes in a stable manner. It implies that at least for our parameter choice the values of QNMs are not significantly affected by the interaction of $\epsilon_{12} = 1$.

\subsection{Superradiance}
\label{sec_SR}
\begin{figure}[h]
\centering
\includegraphics[width=1\linewidth]{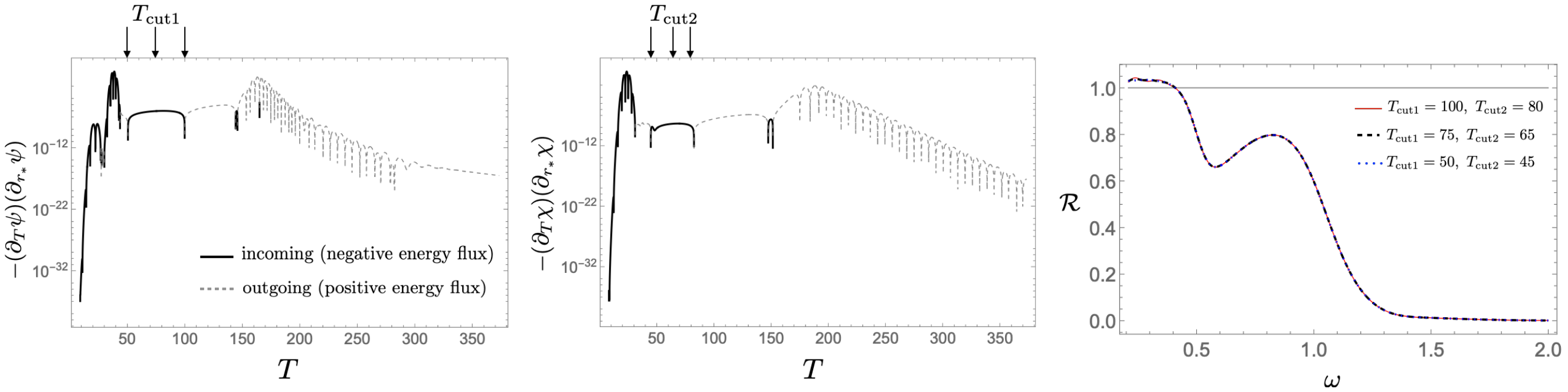}
\caption{{\bf Left panel:} energy flux of $\psi$ measured at $r_{\ast} = 60$ is shown in the time domain and in the log-linear scale. Solid black and dashed gray indicate the incoming and outgoing energy fluxes, respectively. The arrows indicate the three cutoff parameters of $T_{\rm cut 1/cut 2}$ we used in the right panel. We set $\epsilon = 0.5$ and $\epsilon_{12} = 2$, $\epsilon_{11} = \epsilon_{22} = 0$. {\bf Center panel:} energy flux of $\chi$ is shown in the time domain and in the log-linear scale. All used parameters are the same as in the left panel. {\bf Right panel:} reflectivity~${\cal R}$ for the energy flux shown in the left and center panels. For the different three lines, we change the cutoff times~$T_{\rm cut1}$ and $T_{\rm cut2}$ to split the flux into the ingoing and outgoing fluxes in the time domain.
}
\label{pic_enefre}
\end{figure}
We now discuss possible energy extraction in our setup, an exciting possibility of having one field probing the interior of another's horizon. 
In the energy extraction in the Kerr spacetime, the accessibility of the apparent negative energy inside the ergosphere is essential. Even in our model without spin, the region between the two distinct horizons plays the role of the ergosphere in the Kerr solution in the sense that one field cannot access there (which leads to negative-energy modes inside the exterior horizon) but the other field can do and can extract positive-energy modes by leaving negative-energy modes there. It is nothing but the Penrose process.

To numerically confirm this scenario, we compute the reflectivity~${\cal R}$ defined in Eq.~\eqref{reflectivity_dif}. We first decompose the energy flux into two sectors, ingoing and outgoing fluxes, by separating the time-domain data at $T = T_{\rm cut 1/cut 2}$ (see Sec.~\ref{sec_Diagnostics}).
We have a silent phase between the ingoing and outgoing wave packets in the time domain, where the amplitude is quite small.
We set the value of $T_{\rm cut 1/ cut2}$ within the silent phase and when the sign of the energy flux flips from negative (ingoing) to positive (outgoing). During the silent phase, the sign flips several times, which may be caused by numerical errors. However, the resulting reflectivity~${\cal R}$ is insensitive to the choice of $T_{\rm cut}$ as long as it is within the silent domain as shown in the right panel in Fig.~\ref{pic_enefre}.
We then perform the Fourier analysis and obtain the reflectivity ${\cal R}$ as described in Sec.~\ref{sec_Diagnostics}. 

\begin{figure}[t]
\centering
\includegraphics[width=0.45\linewidth]{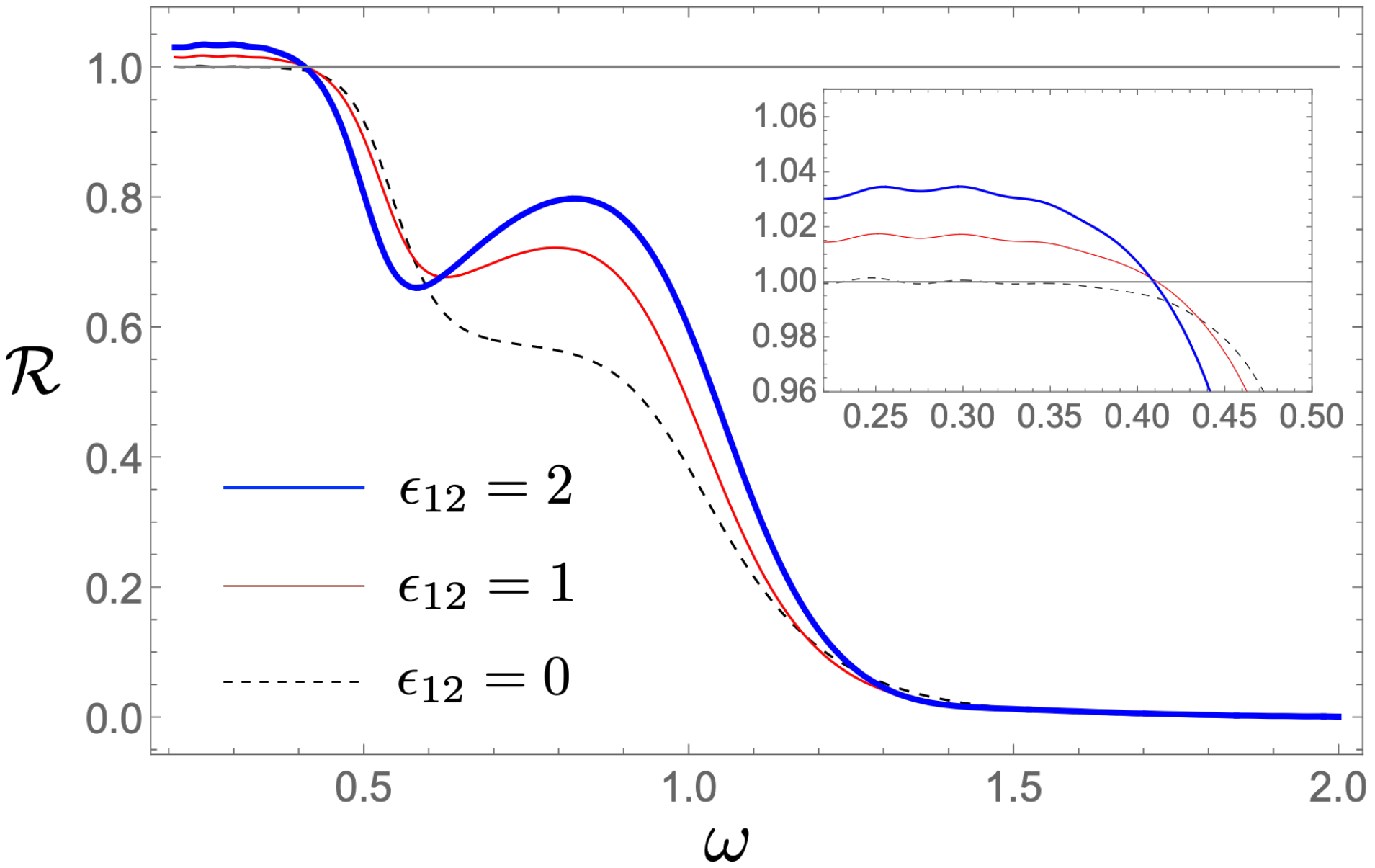}
\caption{Reflectivity ${\cal R}$ with $\epsilon_{11} = \epsilon_{22} = 0$ and $\epsilon = 0.5$. We take $A_{\psi} = 1 = A_{\chi}$. 
}
\label{pic_inte}
\end{figure}

\begin{figure}[h]
\centering
\includegraphics[width=0.45\linewidth]{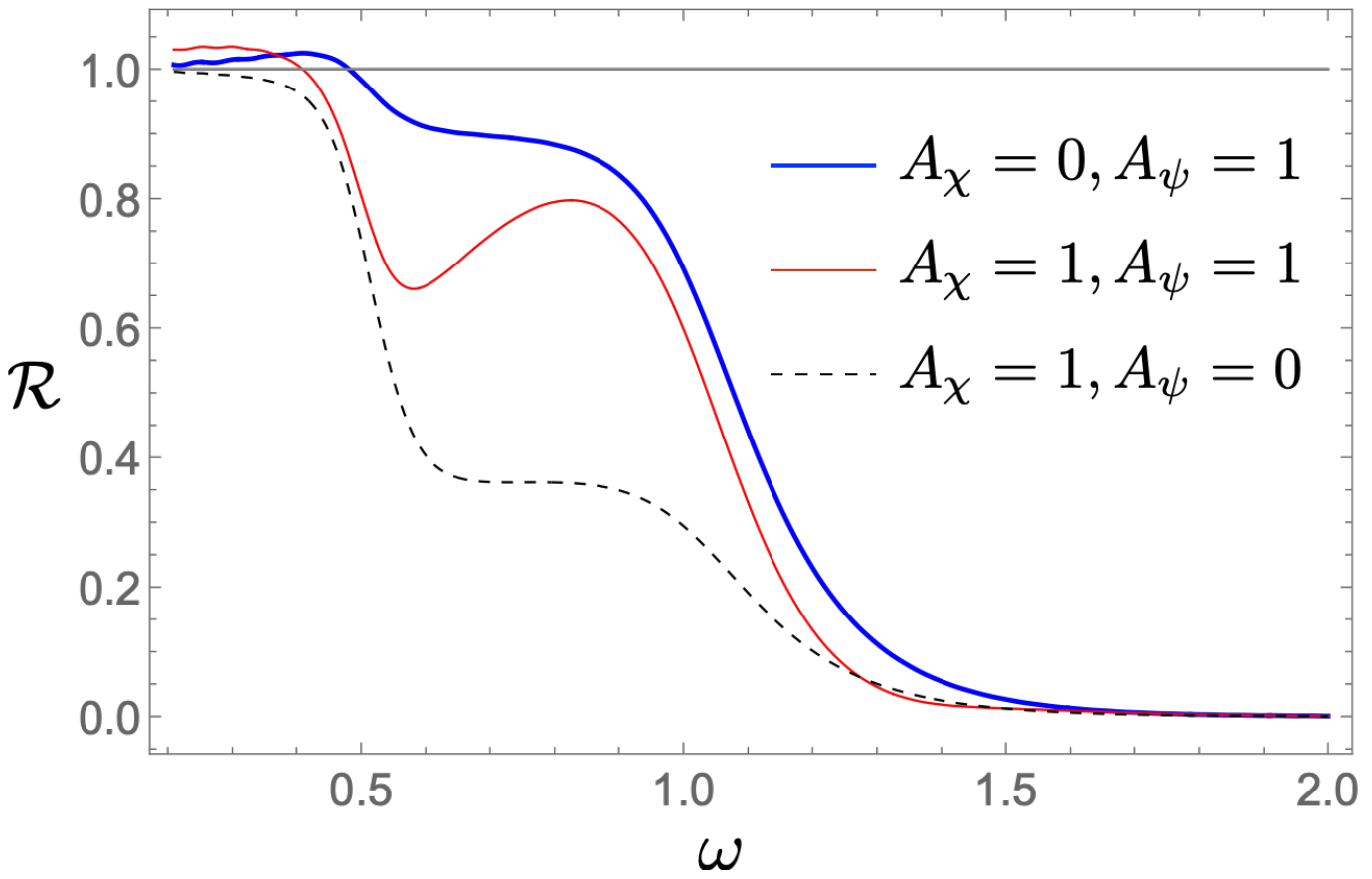}
\caption{Reflectivity ${\cal R}$ with $\epsilon_{11} = \epsilon_{22} = 0$, $\epsilon_{12} = 2$, and $\epsilon = 0.5$.
}
\label{pic_injection}
\end{figure}

\begin{figure}[t]
\centering
\includegraphics[width=0.45\linewidth]{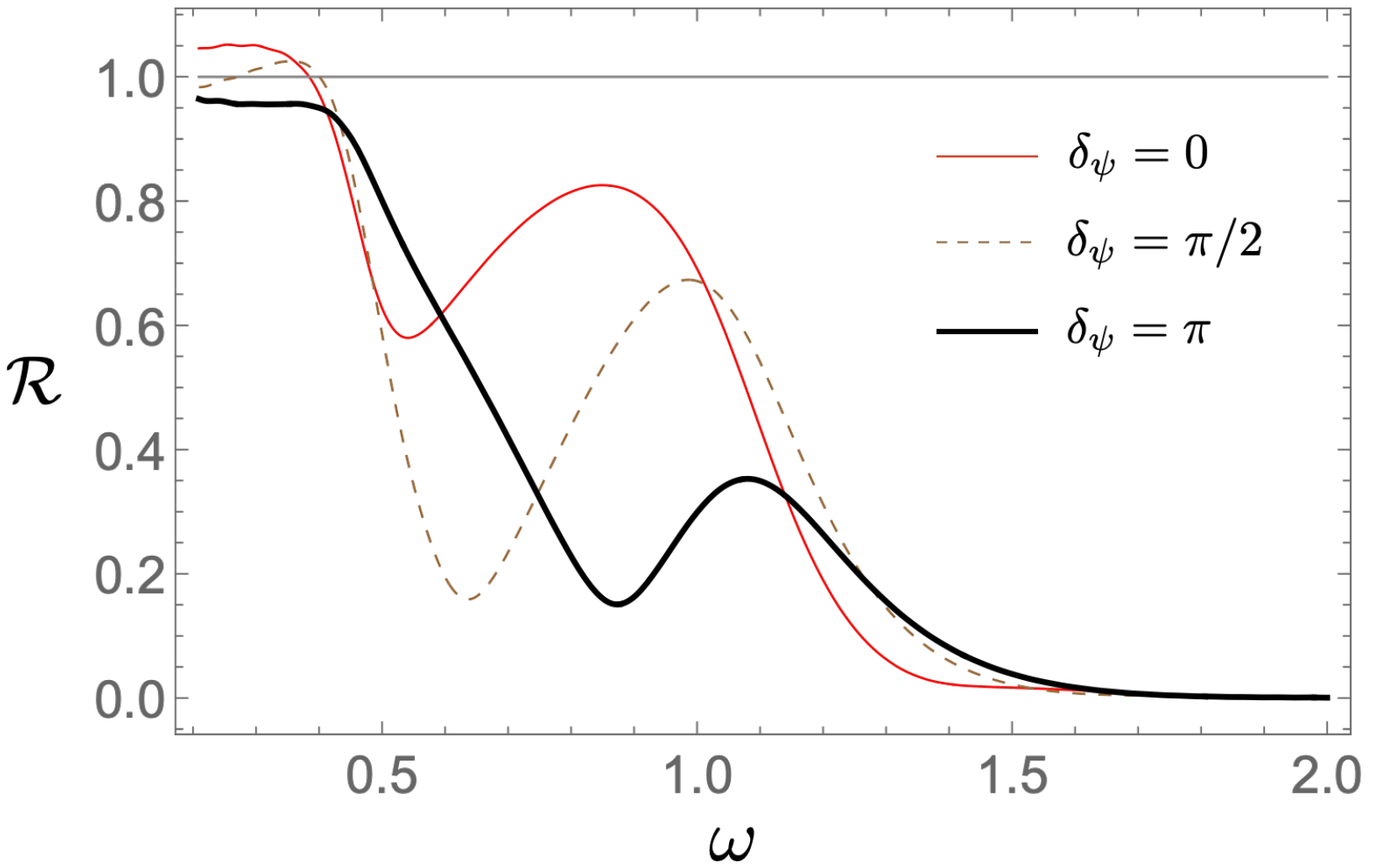}
\caption{Reflectivity for different phases of $\psi$'s wave packet: $\delta_{\psi} = 0$, $\pi/2$, and $\pi$. We set $\epsilon_{11}= \epsilon_{22}=0$, $\epsilon_{12} = 3$, and $\epsilon = 0.5$. Another phase $\delta_{\chi}$ is set to zero and $A_{\psi} = 1 = A_{\chi}$.
}
\label{pic_interference}
\end{figure}

\begin{figure}[t]
\centering
\includegraphics[width=0.45\linewidth]{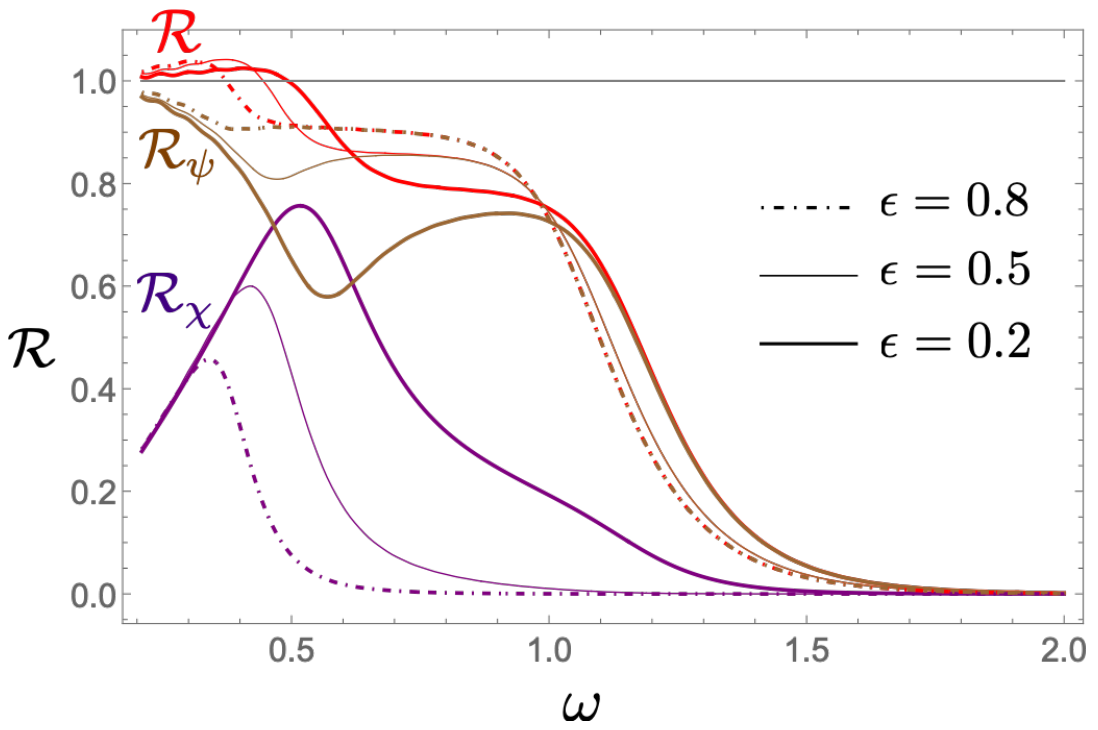}
\caption{Reflectivity ${\cal R}$ with $\epsilon_{11} = \epsilon_{22} = 0$ and $\epsilon_{12} = 3$.}
\label{pic_epsilon}
\end{figure}

\begin{figure}[t]
\centering
\includegraphics[width=0.45\linewidth]{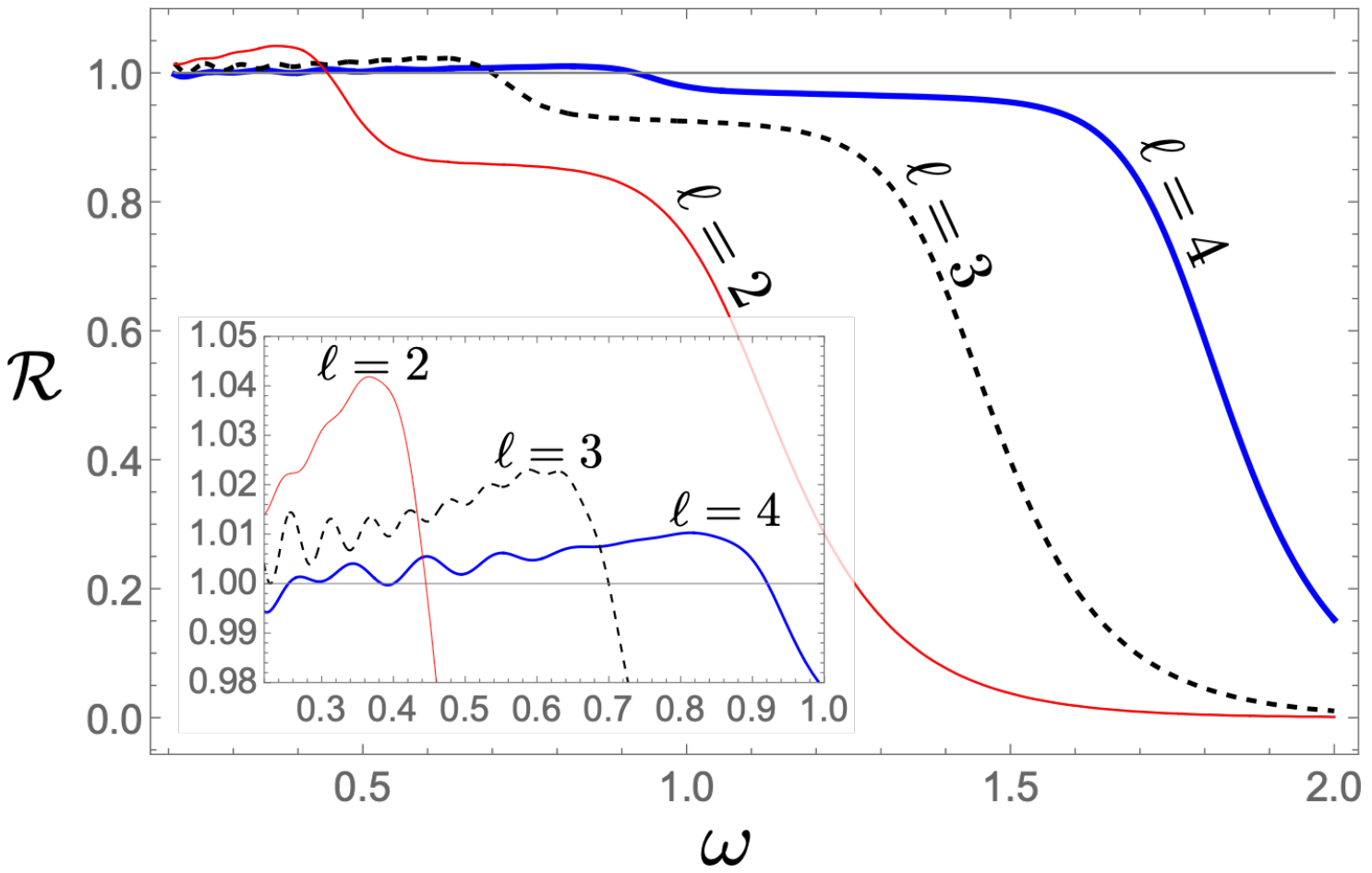}
\caption{Reflectivity ${\cal R}$ for different multipole modes: $\ell =2$ (red thin solid), $\ell =3$ (black dashed), and $\ell =4$ (blue thick solid). We set $\epsilon_{11} = 0 = \epsilon_{22}$, $\epsilon_{12} = 3$, and $\epsilon = 0.5$.}
\label{pic_ell}
\end{figure}

\begin{figure}[h]
\centering
\includegraphics[width=0.95\linewidth]{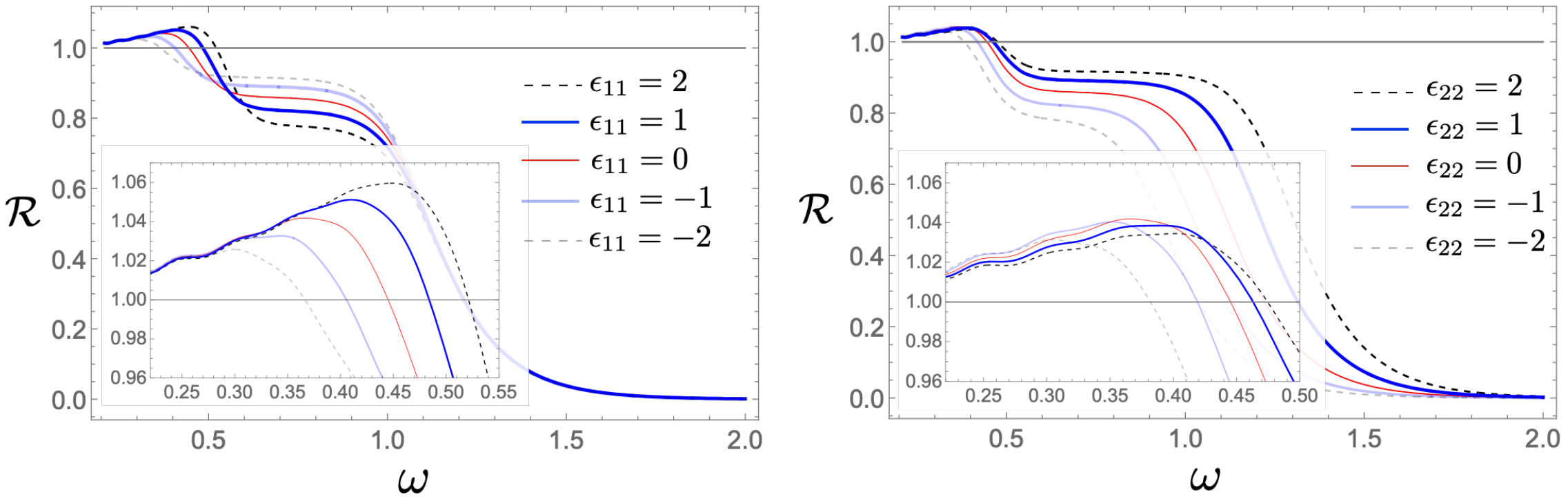}
\caption{Reflectivity ${\cal R}$ (left) for different values of $\epsilon_{11}$ with $\epsilon_{22} =0$ and (right) for different values of $\epsilon_{22}$ with $\epsilon_{11} =0$. We set $\epsilon_{12} = 3$ and $\epsilon = 0.5$.}
\label{pic_massive}
\end{figure}

We find superradiant amplification of waves, or in other words, energy extraction out of a BH, as the reflectivity exceeds unity at lower frequencies. The reflectivity we obtained is insensitive to the choice of the value of $T_{\rm cut 1/cut 2}$. Our results are summarized in Figs.~\ref{pic_enefre}--\ref{pic_massive}. Here, we fix $r^{\psi}_{\ast,s} = 80 \sqrt{1+\epsilon}$ for $\psi$ and $r^{\chi}_{\ast,s} = 80$ for $\chi$ so that the two wave packets arrive simultaneously to the near-horizon region.

Fluxes are shown in Fig.~\ref{pic_enefre}, where we can see that the pulse is initially ingoing, after which the wave packets interact with the geometry and get scattered back to large spatial regions. A clear ringdown at late stages is seen, whose features were already discussed in Sec.~\ref{sec_QNM}. However, the right panel shows that reflectivity~${\cal R}$ [as defined in Eq.~\eqref{reflectivity_dif}] for this set of parameters is larger than unity at small frequencies. The fields are extracting energy away from the geometry, for reasons explained heuristically in the Introduction.
It is interesting that such ``superradiant'' energy extraction takes place at low frequencies only, something which is explained by the fact that high-frequency waves are hard to scatter back and simply fall onto the inner horizon.

%
It is natural to expect that, since there is no energy extraction in the decoupling limit $\epsilon_{12}=0$, energy extraction is larger at larger couplings. This expectation is borne out in our numerical results, Fig.~\ref{pic_inte}. We can also see two strong
suppressions at different frequencies in ${\cal R}$. This is caused by the fact that the two light rings for $\psi$ and $\chi$ lead to an abrupt decrease in the reflectivity at different frequencies as the two light rings have different heights.

Let us now study the dependence of the superradiant amplification on the initial data. Figure~\ref{pic_injection} shows the reflectivity~${\cal R}$ for different combinations of the initial wave amplitudes, $A_\psi$ and $A_\chi$.
The other parameters of the wave packets are set to the same ones used in the ringdown analysis. We consider the following initial data: $(A_{\psi}, A_{\chi}) = (1,1)$, $(A_{\psi}, A_{\chi}) = (1,0)$, and $(A_{\psi}, A_{\chi}) = (0,1)$. We find superradiant amplification when $\psi$'s wave packet that can access the horizon interior is injected. On the other hand, as is apparent in Fig.~\ref{pic_injection}, we observe less significant energy extraction when $A_{\chi} = 0$ and $A_{\psi} = 1$ with $\epsilon>0$.
In a more realistic setup, events that could excite significantly and simultaneously both the observable (e.g., gravitons) and hidden sectors are BH mergers. Our setup~$A_{\psi} =1= A_{\chi}$ could be a mimicker of such scenarios, to see the superradiant amplification that may be caused by strong gravity phenomena.

Given that the two wave packets are simultaneously injected, interference between $\psi$ and $\chi$ affects the superradiant amplification. Figure~\ref{pic_interference} shows the reflectivity~${\cal R}$ for different initial phases of $\psi$'s wave packet. We can infer that the energy extraction due to the multiple speeds of propagation is sensitive not only to the interaction but also to the details of the injected waves, such as interference effects.

Let us now ``freeze'' the interference and study instead the dependence on the horizon size. To this end, we inject a single wave packet of a field that can access the interior of the outer horizon. That is, we consider $(A_{\psi}, A_{\chi}) = (1,0)$. From our result shown in Fig.~\ref{pic_epsilon}, one can read that both the ${\cal R}_{\psi}$ and ${\cal R}_{\chi}$ are less than unity, which means that {\it both} fields are important to yield superradiance. Also, at lower frequencies, the superradiance is significant for a larger separation of the two horizons, i.e., a larger value of $\epsilon$. This is reasonable as a larger region to cause superradiance can efficiently accommodate lower-frequency modes relevant to superradiance.
On the other hand, the frequency range in which superradiance appears is narrower for larger values of $\epsilon$. The height of the angular momentum barrier for $\chi$ is suppressed and it cannot contribute to the reflectivity at higher frequencies of $\omega \gtrsim 1/[r_s (1+\epsilon)]$.

The height of the angular momentum barrier depends on the multipole mode~$\ell$ and the mass of $\psi$ and $\chi$ as well. 
Indeed, we find that the superradiant frequency is larger for a large value of $\ell$ as is shown in Fig.~\ref{pic_ell}.
Figure~\ref{pic_massive} shows that the superradiant frequency increases for massive cases.
For a higher multipole mode or massive modes, higher-frequency modes are scattered by high potential barriers, which increases the superradiant frequency.

\section{Conclusions}\label{sec:conc}
We have shown an explicit realization of a theory of two coupled scalars which probe different effective geometries, and which propagate on distinct spacetimes. Our results are very clear: one field can probe the BH interior of the other via their coupling and transport outward important information on the underlying theory. In particular, the relaxation of BHs in our setup proceeds via {\it two} dominant QNMs, corresponding to the fundamental modes of the two effective geometries. Also, our setup exhibits energy extraction out of a nonspinning BH as expected from the following reasons.

In our system, the field~$\chi$ that sees the outer horizon can indirectly access the horizon interior via the interaction with $\psi$ that probes the inner horizon. Given a field that can access the horizon interior where apparent negative energy is available, it could trigger the Penrose process even without the spin of the BH~\cite{Eling:2007qd,Brito:2015oca,Oshita:2021onq}. We indeed find superradiance in our setup, by introducing multiple speeds of propagation in the BH background. These are exciting results. The superradiance amplification factors are large (i.e., without any tuning of parameters, we find amplification of a few percent). In addition, new phenomena might be possible, such as instabilities, if this system is enclosed in a cavity or if the fields are massive~\cite{Brito:2015oca}.

A classical result concerning ergoregions states that asymptotically flat, horizonless spacetimes with ergoregions are dynamically unstable~\cite{Friedman:1978ygc,Sato:1978ue,Brito:2015oca,Moschidis:2016zjy,Mukohyama:1999kj}.
We can take inspiration from such a result. If we do away with the horizon for $\psi$, for example by filling the interior of $r<r_s(1+\alpha\epsilon)$, with $\alpha<1$ positive, then $\chi$ sees a horizon of ``size''~$r_s(1+\epsilon)$ inside of which it carries negative energies. However, there is no horizon for $\psi$; all it sees is a star. Nevertheless, the field~$\psi$ has access to an ergoregion (that of $\chi$, through the coupling), and one may expect instabilities to develop, for any sign of $\epsilon_{ij}$. A precise study of this phenomena is outside the scope of this work.

We expect that our results would apply to any situation where two (or more) coupled degrees of freedom have different propagation speeds.\footnote{
Although having (spontaneously or explicitly) Lorentz-breaking gravitational theories and starting with the action~(\ref{eqn:action1}) is one of the possibilities leading to such a situation, one could start with Eq.~(\ref{eqn:action2}) without making an assumption on the gravity sector or specifying the relation between $g_{\mu\nu}$ and $\tilde{g}_{\mu\nu}$. When the two metrics do not share the same horizon and the two fields propagating on each metric are interacting with each other, we expect to see results qualitatively similar to what we have presented in the present paper.
Having said this, it is also true that direct interactions between the two matter sectors tend to introduce instabilities such as Boulware-Deser ghost, in particular in the context of massive gravity or bigravity~\cite{Yamashita:2014fga,deRham:2014fha,Gumrukcuoglu:2015nua}.
The model described by the action~(\ref{eqn:action1}) trivially evades this issue and one can safely promote $g_{\mu\nu}$ to a dynamical metric without any fatal instabilities.} Such a situation often happens in modified gravity models where additional field(s) are present on top of the metric in the gravity sector. For instance, in the context of scalar-tensor theories, black hole perturbations have been studied extensively, where the scalar mode does not travel at the same speed as that of the metric perturbations in general (see, e.g., Refs.~\cite{Kobayashi:2014wsa,Franciolini:2018uyq,Khoury:2020aya,Takahashi:2021bml}). This offers an interesting possibility that the energy extraction from a BH and the characteristic late-time relaxation, which we have demonstrated for a simple toy model, could be found in gravitational wave observations. We leave this issue for future work.


\acknowledgments
Vitor Cardoso is grateful for the warm hospitality at YITP.
We acknowledge support by VILLUM Foundation (Grant No.\ VIL37766) and the DNRF Chair program (Grant No.\ DNRF162) by the Danish National Research Foundation.
V.C.\ is a Villum Investigator and a DNRF Chair.  
V.C.\ acknowledges financial support provided under the European Union’s H2020 ERC Advanced Grant “Black holes: gravitational engines of discovery” Grant No.\ Gravitas–101052587. 
This project has received funding from the European Union's Horizon 2020 research and innovation programme under the Marie Sk{\l}odowska-Curie Grant No. 101007855 and No. 101131233.
The work of S.M.~was supported in part by 
Japan Society for the Promotion of Science Grants-in-Aid for Scientific Research Grant No.~24K07017 and the World Premier International Research Center Initiative (WPI), MEXT, Japan. 
The work of N.O.~was supported by Japan Society for the Promotion of Science KAKENHI Grant No. JP23K13111 and by the Hakubi project at Kyoto University.
K.T.~was supported by Japan Society for the Promotion of Science KAKENHI Grants No.~JP22KJ1646 and No. JP23K13101.
Views and opinions expressed are however those of the author only and do not necessarily reflect those of the European Union or the European Research Council. Neither the European Union nor the granting authority can be held responsible for them.


\appendix*

\section{Resolution test and sanity check}
\label{app_resolution}
\begin{figure}[h]
\centering
\includegraphics[width=0.45\linewidth]{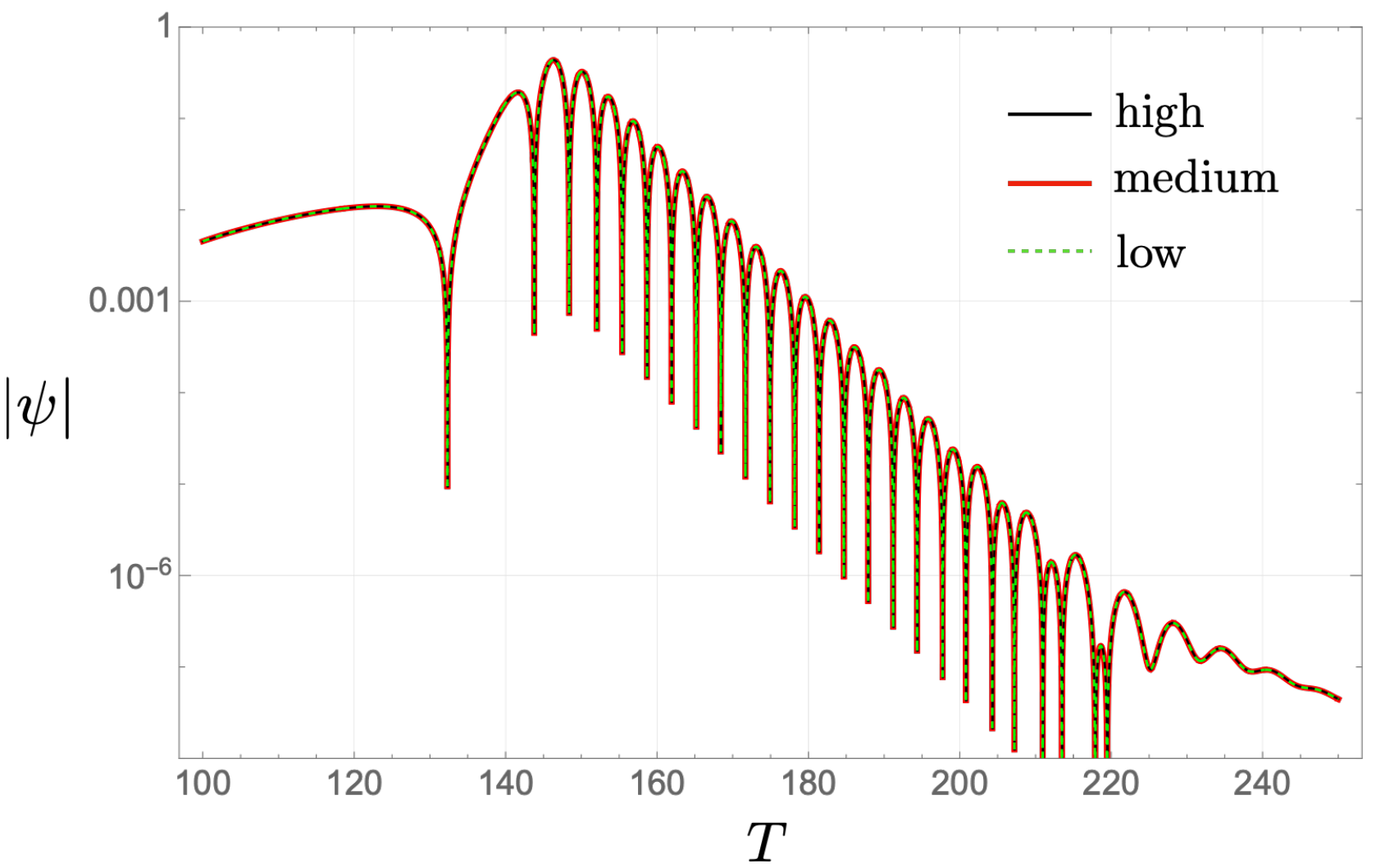}
\caption{Resolution test with $\Delta r_{\ast} = 700/6000$ (low), $700/7500$ (medium), and $700/9000$ (high). The plot is shown in the log-linear scale.}
\label{pic_resolution}
\end{figure}

When we solve the PDEs~\eqref{PDE1}--\eqref{PDE4} numerically, we need to specify the time step~$\Delta T$ and the spatial grid size~$\Delta r_{\ast}$.
Here, we fix the ratio between them as
\begin{equation}
\lambda \coloneqq \frac{\Delta T}{\Delta r_{\ast}} = 0.1\,,
\end{equation}
which is less than unity and satisfies the Courant condition. We then solve the PDEs 
in the range of $-100 \leq r_{\ast} \leq 600$ with $7500$ bins, i.e.,  $\Delta r_{\ast} = 700/7500$. 
We perform the resolution test with three different resolutions as is shown in Fig.~\ref{pic_resolution}.

\begin{figure}[h]
\centering
\includegraphics[width=0.45\linewidth]{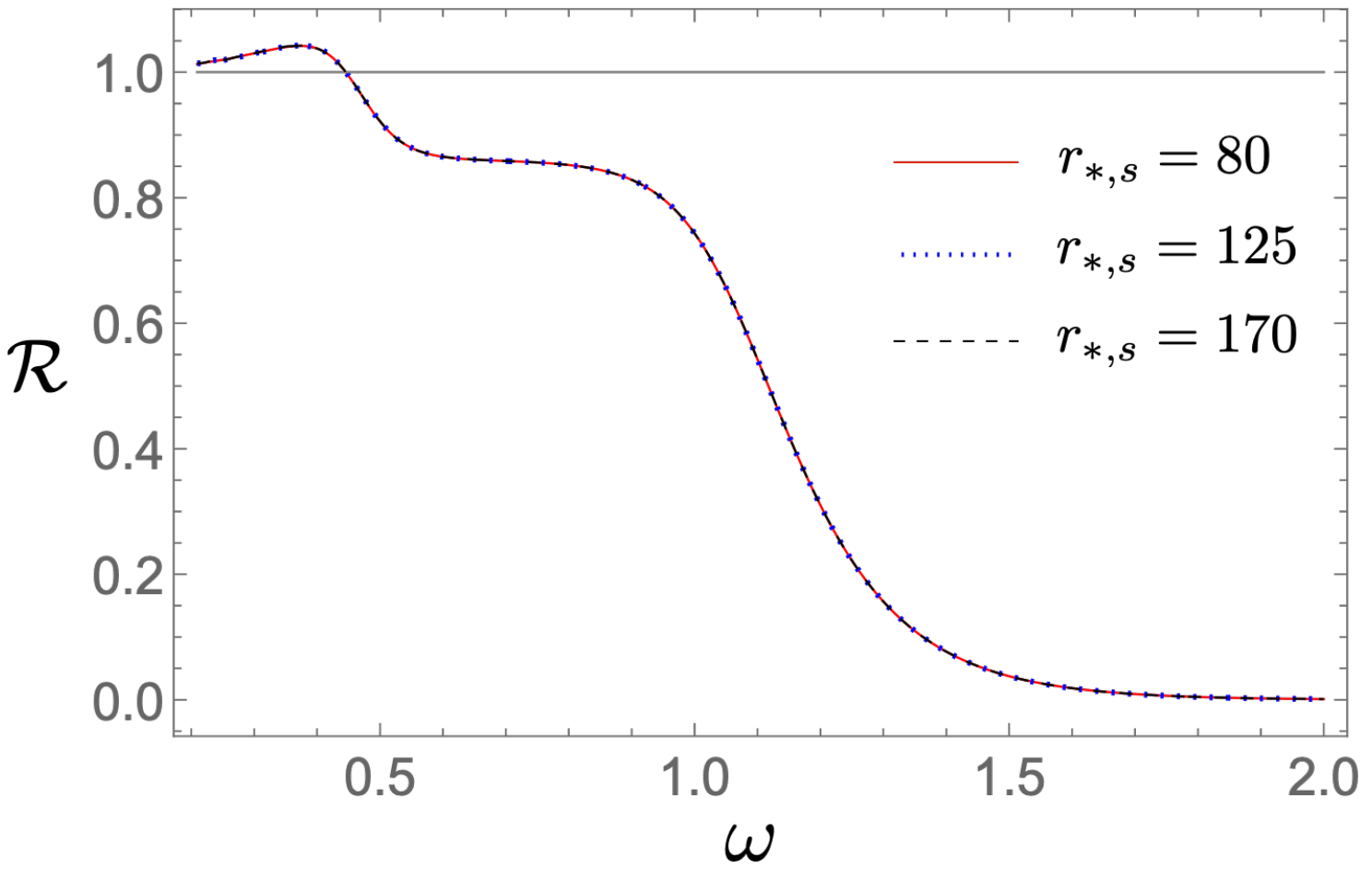}
\caption{Reflectivity~${\cal R}$ for different initial positions of $\psi$'s wave packet: $r_{\ast,s} = 80$, $125$, and $170$. We set here $\epsilon_{11} = \epsilon_{22} = 0$, $\epsilon_{12}=3$, $\epsilon = 0.5$, $A_{\psi}=1$, and $A_{\chi} = 0$. We read the time-domain data at $r_{\ast,o} = 60$.}
\label{pic_distance}
\end{figure}

We also check the superradiant amplification, discussed in Sec.~\ref{sec_SR}, is stable against the position of the initial wave packet. To exclude the interference effect on the superradiance reported in Fig.~\ref{pic_interference}, we inject $\psi$'s wave packet only (see Fig.~\ref{pic_distance}).
%

\bibliographystyle{mybibstyle}

\end{document}